\documentclass[%
reprint,
dvipsnames,
 citeautoscript,
twocolumn,
 amsmath,amssymb,
 longbibliography,
pre,
floatfix,
]{revtex4-2}
\usepackage{lipsum} 

\usepackage{amsmath}
\usepackage{amsthm}
\usepackage{amsfonts}
\usepackage{amssymb}
\usepackage{graphicx}
\usepackage{numprint}
\usepackage{mathtools}
\usepackage{siunitx}
\usepackage{float}
\usepackage{array}
\usepackage{paralist}
\usepackage{cases}
\usepackage{subfigure}
\usepackage{caption}
\usepackage{float}
\usepackage{braket}
\usepackage{mathrsfs}
\usepackage{amsmath,bm}
\usepackage{multirow}
\usepackage[normalem]{ulem}
\usepackage{amsmath} 
\usepackage{amsthm} 
\usepackage{amssymb} 
\usepackage{setspace} 
\usepackage[colorlinks=true,linkcolor=blue,citecolor=blue,urlcolor=cyan,bookmarks,breaklinks=true]{hyperref}
\usepackage{enumitem}
\usepackage{graphicx}
\usepackage{dcolumn}
\usepackage{bm}
\usepackage{tikz}
\usepackage{siunitx}
\usepackage{tabularx} 
\usepackage{parskip}
\usepackage{booktabs}
\usepackage{xcolor}
\usepackage[font=small,labelfont=bf,
   justification=raggedright,
   format=plain]{caption}

\begin{document}
\title{Hyperuniformity Classes of Quasiperiodic Tilings via Diffusion Spreadability}
\begin{abstract} 

Hyperuniform point patterns can be classified by the hyperuniformity scaling exponent $\alpha > 0$, that characterizes the power-law scaling behavior of the structure factor $S(\mathbf{k})$ as a function of wavenumber $k\equiv|\mathbf{k}|$ in the vicinity of the origin, e.g., $S(\mathbf{k})\sim|\mathbf{k}|^{\alpha}$ in cases where $S(\mathbf{k})$ varies continuously with $k$ as $k\rightarrow0$.
In this paper, we show that the spreadability is an effective method for determining $\alpha$ for quasiperiodic systems where $S(\mathbf{k})$ is discontinuous and consists of a dense set of Bragg peaks.
It has been shown in [Torquato, Phys. Rev. E 104, 054102 (2021)] that, for media with finite $\alpha$, the long-time behavior of the excess spreadability $\mathcal{S}(\infty)-\mathcal{S}(t)$ can be fit to a power law of the form $\sim t^{-(d-\alpha)/2}$, where $d$ is the space dimension, to accurately extract $\alpha$ for the continuous case.
We first transform quasiperiodic and limit-periodic point patterns into two-phase media by mapping them onto packings of identical nonoverlapping disks, where {space interior to the} disks represents one phase and the space in {exterior to} them represents the second phase. We then compute the spectral density ${\Tilde{\chi}_{_V}(k)}$ of the packings, and finally compute and fit the long-time behavior of their excess spreadabilities.
Specifically, we show that the excess spreadability can be used to accurately extract $\alpha$ for the 1D limit-periodic period doubling chain ($\alpha = 1$) and the 1D quasicrystalline Fibonacci chain ($\alpha = 3$) to within $0.02\%$ of the analytically known exact results.
Moreover, we obtain a value of $\alpha = 5.97\pm0.06$ for the 2D Penrose tiling, which had not been computed previously,
and present plausible theoretical arguments strongly suggesting that $\alpha$ is exactly equal to 6.
We also show that, due to the self-similarity of the structures examined here, one can truncate the small-$k$ region of the scattering information used to compute the spreadability and obtain an accurate value of $\alpha$, with a small deviation from the untruncated case that decreases as the system size increases.
This strongly suggests that one can obtain a good estimate of $\alpha$ for an infinite self-similar quasicrystal from a modestly-sized finite sample.
The methods described here offer a simple way to characterize the large-scale translational order present in quasicrystalline and limit-periodic media in any space dimension that are self-similar. Moreover, the scattering information extracted from these two-phase media encoded in ${\Tilde{\chi}_{_V}(k)}$, can be used to estimate their physical properties, such as their effective dynamic dielectric constants, effective dynamic elastic constants, and fluid permeabilities.

\end{abstract}
\author{Adam Hitin-Bialus}
\email[Email: ]{hitinbialus@mail.tau.ac.il}
\altaffiliation{Present affiliation: Raymond and Beverly Sackler School of Physics and Astronomy, Tel Aviv University, Tel Aviv, 6997801, Israel}
\affiliation{Department of Physics, Princeton University, Princeton, NJ, 08544, USA}
\author{Charles Emmett Maher}
\email[Email: ]{cemaher@princeton.edu}
\affiliation{Department of Chemistry, Princeton University, Princeton, New Jersey, 08544, USA}
\author{Paul J. Steinhardt}
\email[Email: ]{steinh@princeton.edu}
\affiliation{Jefferson Physical Laboratory, Harvard University, Cambridge, MA 02138 USA}
\affiliation{Department of Physics, Princeton University, Princeton, New Jersey 08544, USA}
\author{Salvatore Torquato}
\email[Email: ]{torquato@princeton.edu}
\affiliation{Department of Chemistry, Princeton University, Princeton, NJ, 08544, USA}
\affiliation{Department of Physics, Princeton University, Princeton, NJ, 08544, USA}
\affiliation{Princeton Institute of Materials, Princeton University, Princeton, NJ, 08544, USA}
\affiliation{Program in Applied and Computational Mathematics, Princeton University, Princeton, NJ, 08544, USA}

\date{\today}
\maketitle
\newpage
\section{Introduction}\label{Intro}
Hyperuniformity generalizes the traditional concept of long-range order and allows for the classification of all perfect crystals, perfect quasicrystals, and special disordered systems with strongly correlated, suppressed long wavelength density fluctuations \cite{hyperuniformityI,TorquatoBook,some_spectral}.
Hyperuniformity is am emerging interdisciplinary field, impinging on a wide range of physical phenomena relevant to photonic and phononic band-gap materials \cite{man1,hyperuniform_properties_2,klatt2022wave, aubry2020experimental,froufe2017band}, antenna and laser design \cite{antenna}, thermal properties of stealthy systems \cite{thermal_properties_1,thermal_properties_2}, transport properties and critical currents in superconductors \cite{transport_1}, diffusion processes in two-phase media \cite{TorquatoDiff,haina,SKOLNICK}, and pure mathematics \cite{primes,puremath2,puremath3,ghosh2018generalized}. 
Past investigations of the novel properties of hyperuniform two-phase media \cite{hyperuniformity_numerical_1} are particularly relevant to this work.

Hyperuniform point configurations (patterns) in $d$-dimensional Euclidean space $\mathbb{R}^d$, possess structure factors $S(\textbf{k})$ with the property
\begin{equation}
         \lim_{|\textbf{k}| \rightarrow 0} S(\textbf{k})  = 0. 
\end{equation}
When the structure factor has a power-law scaling in the vicinity of the origin, i.e. $S(\textbf{k}) \sim |\textbf{k}|^\alpha$, hyperuniform systems are divided into three distinct hyperuniformity classes using the scaling exponent $\alpha$.
Class I ($\alpha>1$) is the ``strongest'' form of hyperuniformity, containing all perfect crystals \cite{hyperuniformityI}, some quasicrystals \cite{some_spectral,quasihyperuniformity,hyper_and_anti}, stealthy and other hyperuniform disordered ground states \cite{hyperuniformity_numerical_2,uche2004constraints}, perturbed lattices and other systems \cite{gabrielli2008tilings,gabrielli2004point}.
Class II ($\alpha=1$) contains some quasicrystals \cite{quasihyperuniformity}, classical disordered ground states \cite{hyperuniformity_numerical_2,zhang2016perfect},
zeros of the Riemann zeta function \cite{torquato2008point,montgomery1973pair}, and a few other systems (see Refs. \citenum{mehta2004random,torquato2008point} and \citenum{feynman1956energy} for more examples). Class III ($0<\alpha<1$) is the ``weakest'' form of hyperuniformity which contains classical disordered ground states \cite{zachary2011anomalous}, random organization models \cite{hexner2015hyperuniformity,tjhung2015hyperuniform}, perfect glasses \cite{zhang2016perfect}, and perturbed lattices \cite{kim2018effect}. 

The concept of hyperuniformity can also be extended to two-phase heterogeneous media, which are hyperuniform when \cite{hyperuniformstatesofmatter, some_spectral, torquato2016disordered} 
\begin{equation}
       \lim_{|\textbf{k}|\rightarrow 0} \Tilde{\chi}_{_V}(\textbf{k}) = 0, 
\end{equation}
where the spectral density $\Tilde{\chi}_{_V}(\textbf{k})$ is the two-phase medium analog to $S(\mathbf{k})$.
When the spectral density scales as a power-law near the origin, i.e. $\Tilde{\chi}_{_V}(\textbf{k}) \sim |\textbf{k}|^\alpha$, one can analogously divide hyperuniform two-phase media into three distinct classes based on $\alpha$ \cite{some_spectral, torquato2016disordered}.

Quasicrystals are a state of matter that possesses long-range orientational order, but exhibits quasiperiodic rather than periodic translational order \cite{ostlund1987physics}. 
Characterizing the hyperuniformity of a quasicrystal based on the behavior of $S(\mathbf{k})$ at small $\mathbf{k}$ is problematic because it is discontinuous, consisting of a dense set of Bragg peaks separated by gaps of arbitrarily small size \cite{quasicrystals_first}.
In Ref. \citenum{quasihyperuniformity}, O\u{g}uz et al. demonstrated that a better approach for extracting $\alpha$ is to use the integrated intensity function given by
\begin{equation}\label{zkdef2}
    Z(k) = s_d\int_0^k S(q) q^{d-1}dq,
\end{equation}
where $s_d = d\pi^{d/2}/\Gamma(1+d/2)$ is the surface area of the $d$-dimensional unit sphere and $S(q)$ is the angular-averaged structure factor.
Specifically, for 1D quasicrystals, $Z(k)$ is bounded by functions of the form $ c_-k^{\alpha+1} \le Z(k) \le c_+k^{\alpha+1}$, where $c_{-/+}$ are constants.
While the constants $c_{-/+}$ can be found exactly for the quasiperiodic point patterns used in Ref. \citenum{quasihyperuniformity}, it is not currently known what these constants are for more general classes of quasiperiodic point patterns.
Without these constants, it is difficult to extract $\alpha$ from $Z(k)$ via direct fitting because it is oscillatory as a function of $\log(k)$ \cite{quasihyperuniformity}.
Therefore, we propose using the diffusion spreadability, which has been shown to accurately extract $\alpha$ from a plethora of different two-phase media \cite{TorquatoDiff,haina, superball,SKOLNICK,Maher_Hyperspheres}.

The recently introduced spreadability concept, developed by Torquato \cite{TorquatoDiff}, serves as a link between time-dependent diffusive processes and the microstructure of heterogeneous media across different length scales \cite{TorquatoDiff,haina,SKOLNICK,superball}.
Consider the mass transfer problem in a two-phase medium as a function of time. Assume that initially solute is distributed uniformly only in phase 2 which occupies a volume fraction $\phi_2$ and absent from phase 1 which occupies a volume fraction $\phi_1$. 
Assume also that both phases have the same diffusion coefficient $D$ at all times.
The time-dependent fraction of total solute that is present in phase 1 is termed the \textit{spreadability} $\mathcal{S}(t)$, as it is a measure of the spreadability of information from phase 2 to phase 1.
Torquato expanded the original work done in $\mathbb{R}^3$ by Prager \cite{prager} and showed that in any dimension $d$, the spreadability is related to the microstructure in direct space through the autocovariance function $\chi_{_V}(\textbf{r})$ (defined in Sec. II C) or in Fourier space via the spectral density $\Tilde{\chi}_{_V}(\textbf{k})$ \cite{TorquatoDiff}:
    \begin{equation}\label{spread_twooptions}
    \begin{aligned}
    \mathcal{S}(\infty) - \mathcal{S}(t) &=  \frac{1}{(4\pi D t)^{d/2}\phi_2}\int_{\mathbb{R}^d}  \chi_{_V}(\textbf{r})e^{-r^2 / 4Dt}d\textbf{r}\\
    &=  \frac{1}{(2\pi )^{d}\phi_2}\int_{\mathbb{R}^d}  \Tilde{\chi}_{_V}(\textbf{k})e^{-k^2Dt}d\textbf{k}.
    \end{aligned}
    \end{equation}
Here $\mathcal{S}(\infty)= \phi_1$, and $\mathcal{S}(\infty) - \mathcal{S}(t)$ is called the \textit{excess spreadability}.

\begin{figure*}[!t]
    \centering
    \subfigure[]{\includegraphics[height=0.5\textheight]{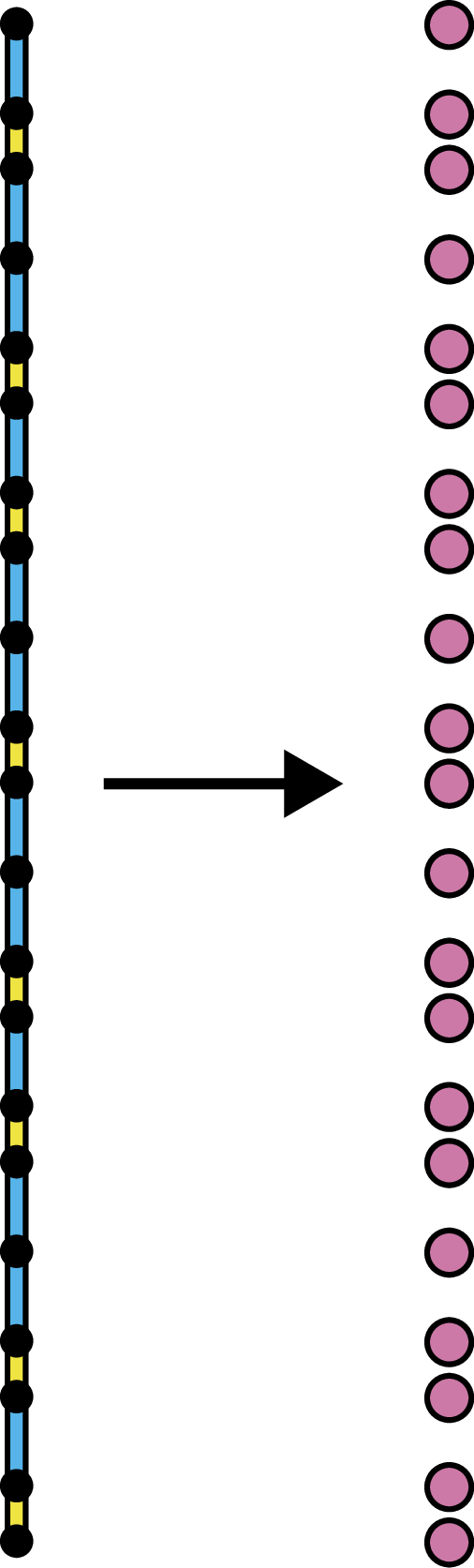} }\\
    \subfigure[]{\includegraphics[width=0.9\textwidth]{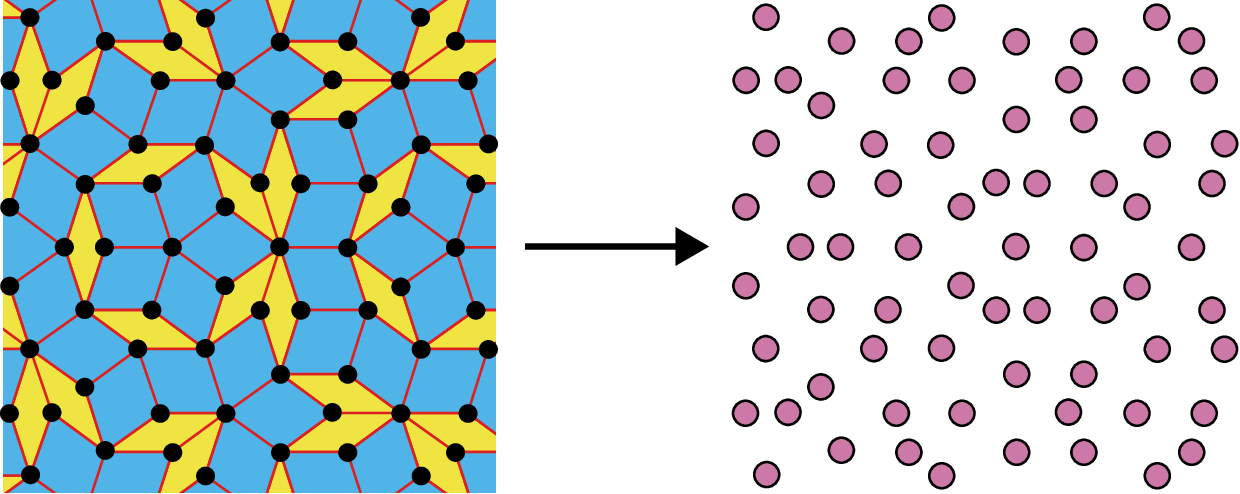} }
     \caption{An illustration of the mapping of vertices of quasicrystalline point configurations (black circles at left) to two-phase media (at right), where the white background is phase 1 and the pink (gray) disks are phase 2. The points for the (a) Fibonacci chain are the endpoints of the long and short segments that compose the chain and the points for the (b) portion of a Penrose tiling and the vertices of the acute and obtuse rhombi that compose the tiling.\label{twophasequasi}}
\end{figure*}

Torquato demonstrated that the small-, intermediate-, and long-time behaviors of $\mathcal{S}(t)$ are directly determined by the small-, intermediate-, and large-scale structural features of the material \cite{TorquatoDiff}.
Specifically, in instances where the spectral density behaves as a power law near the origin, $ \lim_{|\textbf{k}|\rightarrow 0} \Tilde{\chi}_{_V}(\textbf{k}) \sim |\textbf{k}|^\alpha$, the long time excess spreadability scales as $\sim t^{-(d+\alpha)/2}$. 
Hence, the spreadability serves as a dynamical tool to probe and categorize all translationally invariant two-phase media using their long-time scaling, including periodic media \cite{TorquatoDiff}.

The spreadability has been used to classify systems of any hyperuniformity class, including ordered and disordered stealthy systems ($\alpha = \infty$), nonhyperuniform systems ($\alpha = 0$), and antihyperuniform systems $(-d < \alpha < 0)$ \cite{superball,SKOLNICK,haina,TorquatoDiff, Maher_Hyperspheres}. 
Moreover, the spreadability can be measured experimentally and used to quantify the hyperuniformity of a two-phase medium when accurate scattering data near $|\mathbf{k}|=0$ is not available \cite{SKOLNICK}.
Lastly, note that the spreadability involves a Gaussian smoothing of either the autocovariance or spectral density [see Eq. (\ref{spread_twooptions})], which is particularly useful for characterizing systems in which $S(\textbf{k})$ is discontinuous and consists of a dense set of Bragg peaks in the vicinity of the origin. 

The ability of the spreadability to probe systems of different hyperuniformity classes, and perform a smoothing of the spectral density motivates us to extract $\alpha$ from quasiperiodic point configurations in the special case of two-phase media in which the points are decorated with identical nonoverlapping $d$-dimensional spheres, i.e., to create corresponding {\it sphere packings}, where {the space interior to the} spheres represents one phase and the space {exterior to} them represents the second phase.
This particular decoration scheme endows the resulting two-phase medium with the same $\alpha$ as the underlying point pattern (details in Sec. II C).
Specifically, we produce such decorations of periodic approximants of one- and two-dimensional quasiperiodic patterns (cf. Fig. \ref{twophasequasi}) and use the excess spreadability to accurately extract their hyperuniformity exponents $\alpha$.
We first benchmark the spreadability using media with discontinuous $\Tilde{\chi}_{_V}(\textbf{k})$ for which $\alpha$ is known theoretically, namely, those derived from the period-doubling and Fibonacci chains ($\alpha = 1$ and $3$, respectively).
Then, we use the spreadability to measure the value of $\alpha$ of a two-dimensional Penrose quasicrystal point pattern, whose points are the vertices of a Penrose tiling composed of acute and obtuse rhombi (see Sec. \ref{Methods}).

Extracting $\alpha$ from the excess spreadabilities of media with quasiperiodic order requires a more nuanced fitting scheme than the one presented in Ref. \citenum{haina}; see Sec. \ref{Methods}.
We find that the ensemble-averaged values of $\alpha$, extracted from a collection of periodic approximants of infinite quasicrsytals, increase monotonically as the system size increases, and subsequently employ a finite-size scaling analysis to predict the value of $\alpha$ in the thermodynamic limit.
We confirm that this scheme is able to accurately extract the hyperuniformity exponent $\alpha$ from the 1D two-phase media derived from the Fibonacci and period-doubling chains with less than 0.02\% error when compared to known analytic results. 
We then use this scheme on excess spreadability the 2D Penrose tiling-derived medium. 
The value of $\alpha$ has not been computed previously for a Penrose tiling; here we report a value of $\alpha = 5.97 \pm 0.06$ (see Sec. \ref{PT}).

Moreover, by leveraging the self-similarity of the structures considered herein, we demonstrate that one can truncate the small-$k$ region of the spectral density and still extract an accurate value of $\alpha$, with a small deviation from that of the untruncated spectral density that decreases as the system size increases.
Because one can obtain an accurate $\alpha$ by only considering length scales that are a small fraction of the overall system size, this strongly suggests that one can obtain a good estimate of $\alpha$ for an infinite quasicrystal or limit-periodic system by examining a finite quasicrystal or limit-periodic system that is significantly smaller than the largest ones examined in this work, so long as the structure is self-similar.
This is of practical and computational importance because it means one need not examine large quasicrystalline samples, or generate large quasiperiodic point patterns, which are computationally expensive to produce.

This paper is organized as follows:
Section \ref{Background} provides mathematical definitions and preliminaries for hyperuniformity and spreadability. Section \ref{Methods}, discusses the one- and two-dimensional models used in the study, their construction, and the procedures used to extract $\alpha$ from the excess spreadability. Some exact spreadability results are recounted in Sec. \ref{general}. The values of $\alpha$ extracted from the excess spreadability are reported for the period-doubling chain, Fibonacci chain, and Penrose tiling in Secs. \ref{PDC}, \ref{FC}, and \ref{PT}, respectively. Section \ref{Discussion} provides conclusions and directions for further work.

\section{Definitions and preliminaries\label{Background}}

\subsection{Structure factor and number variance in hyperuniform point patterns}
A system of point particles in $\mathbb{R}^d$ can be completely 
characterized by a set of probability functions $\rho_n(\textbf{r}_1,\textbf{r}_2,\ldots,\textbf{r}_n)$ that are proportional to the probability of finding $n$ particles in positions $\textbf{r}_1,\textbf{r}_2,\ldots,\textbf{r}_n$.
For statistically homogeneous systems, the pair correlation function $g_2(\textbf{r}_{12})$ is defined as $g_2(\textbf{r}_{12}) = \rho_2(\textbf{r}_{12})/{\rho^2},$
where $\textbf{r}_{12} = \textbf{r}_{2}-\textbf{r}_{1}$, and $\rho = \rho_1(\textbf{r}_{1})$ is the number density.
The  ensemble-averaged \textit{structure factor} $S(\textbf{k})$ can be defined as: 
\begin{equation}\label{StructureFactor}
    S(\textbf{k}) = 1 + \rho \Tilde{h}(\textbf{k}),
\end{equation}
where $\Tilde{h}(\textbf{k})$ is the Fourier transform of the total correlation function $h(\textbf{r}_{12}) = g_2(\textbf{r}_{12}) - 1$. 
For a single periodic configuration of points, $S(\textbf{k})$ is \cite{hyperuniformstatesofmatter}:
\begin{equation}\label{scatinte}
    S(\textbf{k}) = \frac{| \sum_{j=1}^N e^{-i\textbf{k}\cdot \textbf{r}_j}|^2}{N},\quad \mathbf{k}\neq\mathbf{0}.
\end{equation}
The integrated intensity function $Z(k)$ given in Eq. (\ref{zkdef2}) is an integral over the structure factor, and so tends to smooth over the dense distributions of Bragg peaks and other discontinuities that $S(k)$ may include \cite{quasihyperuniformity}.

A hyperuniform point pattern is one in which the number variance $\sigma_N^2(R) \equiv \braket{N(R)^2} - \braket{N(R)}^2$ of particles in a spherical observation window of radius $R$ grows more slowly than the the window volume in the large-$R$ limit (i.e., more slowly than $R^d$). For a general translationally invariant point configuration in $\mathbb{R}^d$, the local number variance can be expressed exactly in Fourier space using the structure factor \cite{hyperuniformityI}: 
\begin{equation}\label{num_var_with av_sk}
\sigma_N^2(\textbf{R}) = \braket{N(\textbf{R})} \left[ 
\frac{1}{(2\pi)^d} \int_{\mathbb{R}^d} 
 S(\textbf{k}) \Tilde{\alpha}_2(\textbf{k};\textbf{R})d\textbf{k}\right].
\end{equation}
The number variance can also be written in terms of the integrated intensity function \cite{quasihyperuniformity}:
\begin{equation}\label{num_var_from_zk}
    \sigma_N^2(R) = -\rho v_1 (R) \left[ 
    \frac{1}{(2\pi)^d} \int_{0}^{\infty} 
    Z(k) 
  \frac{\partial \Tilde{\alpha}_2(k;R)}{\partial k}dk\right].
\end{equation}
Here, $\Tilde{\alpha}_2(\textbf{k};\textbf{R})$ is the Fourier transform of 
\begin{equation}
    \alpha_2(\textbf{r};\textbf{R}) = \frac{v_2^{\text{int}}(\textbf{r};\textbf{R})}{v_1(R)},
\end{equation}
which is called the scaled intersection volume function, where   
\begin{equation}
    v_2^{\text{int}}(\textbf{r};\textbf{R}) = \int_{\mathbb{R}}w(\textbf{r}+\textbf{x}_0;\textbf{R})w(\textbf{x}_0;\textbf{R})d\textbf{x}_0
\end{equation}
is the intersection volume of two windows with identical orientations whose centers are separated by a displacement vector \textbf{r}. 
In the case of spherical windows, the function $v_1(R)$ is given as the volume of a $d$-dimensional spherical window of radius $R$:
\begin{equation}\label{volume}
    v_1(R) = \frac{\pi^{d/2}}{\Gamma(1+d/2)}R^d.
\end{equation}
It is possible to write the Fourier transform of $ \alpha_2(\textbf{r};\textbf{R})$ as 
\begin{equation}\label{F_alpha_1d}
    \Tilde{\alpha}_2(k,R) = 2^d \pi^{d/2}\Gamma(1+d/2) \frac{[J_{d/2}(kR)]^2}{k^d},
\end{equation}
where $J_{d/2}(kR)$ is the Bessel function of the first kind. From equations (\ref{num_var_with av_sk}) and (\ref{num_var_from_zk}), it follows \cite{hyperuniformityI,zachary2011anomalous} that when the structure factor goes to $0$ continuously at the origin, i.e., $S(\textbf{k)}\sim |\textbf{k}|^\alpha$, hyperuniform systems are divided into three distinct hyperuniformity classes, and the number variance asymptotically scales as:
\begin{numcases}{\sigma_N^2(R) \sim }
    R^{d-1}, \quad \alpha >1  & \(  (\text{Class I})\) \nonumber \\
    R^{d-1}\ln(R),\quad \alpha = 1 & \( (\text{Class II})\)   \nonumber\\
    R^{d-\alpha}, \quad (0 < \alpha < 1), & \( (\text{Class III}).\)   \nonumber
\end{numcases}

\subsection{Hyperuniformity in two-phase media}
A two-phase medium is a partition of space into two disjoint regions called phases \cite{TorquatoBook}. Let phase one occupy a volume fraction $\phi_1$ and phase two occupy a volume fraction $\phi_2 = 1-\phi_1$. 
Here, phase 2 is the packing of identical, nonoverlapping disks (or rods), so $\phi_2$ is also called the {\it packing fraction}.
The two-phase medium can be fully statistically characterized by the $n$-point correlation functions
\begin{equation}\label{Sfunc}
    S_n^{(i)}(\textbf{x}_1,\ldots,\textbf{x}_n) \equiv \braket{\mathcal{I}^{(i)}(\textbf{x}_1)\ldots \mathcal{I}^{(i)}(\textbf{x}_n)}, 
\end{equation}
\noindent where $\mathcal{I}^{(i)}(\textbf{x})$ is the indicator function of phase $i = 1,2$, and the angular brackets indicate an ensemble average. The function $S_n^{(i)}(\textbf{x}_1,\ldots,\textbf{x}_n)$ gives the probability of finding the vectors $\textbf{x}_1,\ldots,\textbf{x}_n$ all in phase $i$. The autocovariance function $\chi_{_V} (\textbf{r})$ is related to the two-point correlation function $S_2^{(i)}(\mathbf{r})$ by
\begin{equation} \label{auto}
    \chi_{_V}(\textbf{r})\equiv S_2^{(1)}(\textbf{r}) - \phi_1^2 = S_2^{(2)}(\textbf{r}) - \phi_2^2 
\end{equation}
assuming statistical homogeneity. 
A useful length scale associated with two-phase media is the specific surface $s$, which is the expected interfacial area between the two phases per unit volume.
For sphere packings, $s$ is given by $d\cdot\phi_2/a$, where $a$ is the sphere radius \cite{torquato1984microstructure}.

Following Ref. \citenum{torquato2016disordered}, for statistically homogeneous and isotropic two-phase media, the autocovariance function $\chi_{_V}(r)$, which depends only on $r\equiv|\mathbf{r}|$, has the asymptotic form \cite{TorquatoBook},
\begin{equation}
    \chi_{_V}(r)=\phi_1\phi_2-\beta(d)sr+\mathcal{O}(r^2),
\end{equation}
in the vicinity of the origin, where
\begin{equation}
    \beta(d)=\frac{\Gamma(d/2)}{2\sqrt{\pi}\Gamma((d+1)/2)}.
\end{equation}
As a consequence, the large-$k$ decay of the corresponding spectral density is controlled by the exact following power-law form:
\begin{equation}
    \Tilde{\chi_{_V}}(k)\sim\frac{\gamma(d)s}{k^{d+1}}, k\rightarrow\infty,
\end{equation}
where
\begin{equation}
    \gamma(d)=2^d\pi^{(d-1)/2}\Gamma((d+1)/2).
\end{equation}

The local volume-fraction variance, $\sigma_{V}^2(R)$, can be written in terms of $\chi_{_V}(\textbf{r})$  as \cite{lu1990local,some_spectral}:
\begin{equation}\label{volumefractionreal}
    \sigma_{V}^2(R) = \frac{1}{v(R)} \int_{\mathbb{R}^d}  \chi_{_V}(\textbf{r})\alpha_2(r;R)d\textbf{r}.
\end{equation}
Via Parseval’s theorem and Eq.(\ref{volumefractionreal}), the volume-fraction
variance has the following Fourier-space
representation:
\begin{equation}\label{volume_fromspec}
    \sigma_{V}^2(R) = \frac{1}{v(R)(2\pi)^d} \int_{\mathbb{R}^d}  
\Tilde{\chi}_{_{V}}(\textbf{k}) \Tilde{\alpha}_2(k;R)d\textbf{k}.
\end{equation}
\indent Following \cite{Torquatotowphase}, a hyperuniform two-phase medium is one with spectral density that vanishes at the origin:
\begin{equation}
    \lim_{|\textbf{k}| \rightarrow 0} \Tilde{\chi}_{_V}(\textbf{k}) = 0.
\end{equation}
As with point configurations, when the spectral density behaves as a power law near the origin, i.e., $\lim_{|\textbf{k}| \rightarrow 0} \Tilde{\chi}_{_V}(\textbf{k}) \approx |\textbf{k}|^\alpha$,
the volume fraction variance of a hyperuniform two-phase medium scales as:
\begin{numcases}{\sigma_V^2(R) \sim }\label{V_classesI}
  R^{-(d+1)}, \quad \alpha >1  & \(  (\text{Class I})\) \nonumber \\
  R^{-(d+1)}\ln(R),\quad \alpha = 1 & \( (\text{Class II})\)   \nonumber\\
  R^{-(d+\alpha)}, \quad (0 < \alpha < 1), & \( (\text{Class III}).\)   \nonumber
\end{numcases}

For packings of identical spheres, the spectral density is exactly given by the product of the structure factor, $\Tilde{\alpha}_2(k;a)$, and packing fraction \cite{TorquatoBook,torquato1985microstructure,torquato2016disordered}, i.e.,
\begin{equation}\label{spectral_nonoverlap}
    \Tilde{\chi}_{_V}(\textbf{k}) = \phi_2 \Tilde{\alpha}_2(k;a)S(\textbf{k}).
\end{equation}
Because $\Tilde{\alpha}_2(k;a)$ scales as a constant as $|\textbf{k}| \rightarrow 0$, Eq. (\ref{spectral_nonoverlap}) implies that the scaling exponent $\alpha$ of $\Tilde{\chi}_{_V}(\textbf{k})$ will be the same as the structure factor $S(\textbf{k})$. 
This does not generally apply to packings of polydisperse spheres, or packings of nonspherical particles \cite{torquato2016disordered}.

We note here that $\sigma_V^2(R)$ and $\sigma_N^2(R)$ cannot be used to extract a precise value of $\alpha$ from class I systems because there is a degeneracy: the number (volume-fraction) variance scales the same way for all class I point patterns (media).
For example, such functions are unable to extract precise values of $\alpha$ for the Fibonacci ($\alpha = 3$) or Penrose ($\alpha \approx 6$, see Sec. \ref{PT}) quasicrystals.
It is therefore necessary to directly fit the small-$k$ behavior of $S(k)$ ($\Tilde{\chi_{_V}}(k)$), or employ a smoothing method such as $Z(k)$ or the spreadability (see below) to determine $\alpha$ for class I point patterns or media.

\subsection{Spreadability}
Torquato \cite{TorquatoDiff} showed that the excess spreadability defined via Eq. (\ref{spread_twooptions}) can be written in terms of angular-averaged autocovariance function $\chi_{_V}(r)$ as follows:
\begin{equation}\label{fourierspread}
\begin{aligned}
    \mathcal{S}(\infty) - \mathcal{S}(t) &= \frac{d\omega_d}{(4\pi D t)^{d/2}\phi_2}\int_0^\infty r^{d-1}  \chi_{_V}(r)e^{-r^2 / 4Dt}dr,
\end{aligned}
\end{equation}
where $\omega_d$ is the volume of a $d$-dimensional sphere of unit radius.
Similarly, in Fourier space, the excess spreadability can be written in terms of the angular-averaged spectral density $\Tilde{\chi}_{_V}(k)$\cite{TorquatoDiff}:
\begin{equation}
\begin{aligned}\label{spreadfromspectral}
    \mathcal{S}(\infty) - \mathcal{S}(t) &= \frac{d\omega_d}{(2\pi)^{d}\phi_2}\int_0^\infty k^{d-1}\Tilde{\chi}_{_V}(k)e^{-k^2Dt}dk. 
\end{aligned}
\end{equation}
Note that the Gaussian kernel in Eq. (\ref{spreadfromspectral}) can be regarded as a smoothing of $\Tilde{\chi}_{_V}(k)$, which we expect to effectively smooth the high-frequency variations of the $\Tilde{\chi}_{_V}(k)$ of the two-phase media considered in this work, which are highly discontinuous in the vicinity of the origin.

Torquato also showed that for two-phase media where the spectral density scales as a power law in the vicinity of the origin, i.e.,
\begin{equation}
    \lim_{|\textbf{k}| \rightarrow 0} \Tilde{\chi}_{_V}(\textbf{k}) \approx |\textbf{k}|^{\alpha}, 
\end{equation}
the long-time behavior of the excess spreadability can be written as \cite{TorquatoDiff}
\begin{equation}\label{longtime}
    \begin{aligned}
     \mathcal{S}(\infty) - \mathcal{S}(t) \sim 1/t^{(d+\alpha)/2}. 
    \end{aligned}
\end{equation}
Using this last equation, it is then possible to extract $\alpha$ from the long time behavior of the spreadability, and so to probe the microstructure of the media.


\section{Methods\label{Methods}}
\subsection{Generating point configurations and decoration of points}
\indent Following O\u{g}uz et al. \cite{hyper_and_anti}, the 1D Fibonacci chain was generated via substitution rules. Starting from a single seed ``link'', a set of substitutions is then applied iteratively to replace it with sets of other links. For example, the Fibonacci chain is a particular case of the substitution rule tiling, constructed by repeated iterations of a substitution of long $L$ and short $S$ links such that $S \rightarrow L$, and $L \rightarrow LS$, where $L$ is of length $\tau = \frac{1+\sqrt{5}}{2}$ and $S$ is of length 1 (see Table \ref{construct}). 
\begin{table}[h!]

\begin{tabular}{|c|c| }
 \hline
  \# of substitutions & Chain\\
 \hline
 $0$ & $S$    \\
 $1$ & $L$  \\
 $2$ & $LS$ \\
 $3$ & $LSL$ \\
 $4$ & $LSLLS$\\
 $5$ & $LSLLSLSL$\\
\hline
\end{tabular}
\caption{Construction of the Fibonacci chain out of consecutive substitutions of the type $S \rightarrow L$, and $L \rightarrow LS$\label{construct}}
\label{Fib subs example}
\end{table}
This kind of substitution can be characterized by a substitution matrix \textbf{M} given by:
\begin{equation}
    \textbf{M} =
    \begin{pmatrix}
    0 & 1 \\
    1 & 1 
    \end{pmatrix},
\end{equation}
which gives the number of $S's$ and $L's$ after a substitution by acting on the two-dimensional column vector $(N_S,N_L)$ which represents the number of $S's$ and $L's$ of the current iteration. A substitution rule for two link types is characterized by the substitution matrix
\begin{equation}
    \textbf{M} =
    \begin{pmatrix}
    a & b \\
    c & d 
    \end{pmatrix},
\end{equation}
where $a,b,c,d$ are integers. O\u{g}uz et al. \cite{hyper_and_anti} conjectured a closed form formula for the hyperuniformity scaling exponent $\alpha$ given by the eigenvalues of the substitution matrix: 
\begin{equation}\label{theory_alpha}
\alpha = 1-2\frac{\ln{|\lambda_2|}}{\ln{|\lambda_1|}},   
\end{equation}
with $\lambda_1 > \lambda_2$. 
In this study, Fibonacci chains with between $N = 2584$ and $N = 514229$ links (particles) were considered, which are class I hyperuniform point patterns with $\alpha = 3$.
Similarly, the limit-periodic period doubling chain can be produced via the substitution rules $L\rightarrow LSS$ and $S \rightarrow L$, where $L$ is length 2 and $S$ is length one, starting from a seed link $S$ \cite{hyper_and_anti}.
In this study we also consider period doubling chains with between $N = 21845$ and $N = 22369621$ links, which are class II hyperuniform with $\alpha = 1$.

The 2D Penrose tiling periodic approximants were created using the Generalized Dual Method, which is a mapping from regions created by infinite intersections of straight lines to points \cite{lin2017hyperuniformity,dealgebraic,socolar1985quasicrystals}.
A periodic grid is an infinite set of parallel, equally spaced, straight lines --- the spacing between lines here is 1. 
One can then label each line by $m\in \mathbb{Z}$ according to its ordinal position in the grid.
Five of these grids are then stacked one atop the other with the $i$th grid oriented normal to: 
\begin{equation}
    \begin{split}
        \hat{\mathbf{r}}_0 &= (1,0), \hat{\mathbf{r}}_1 = (\mathrm{cos}[2\pi/5], \mathrm{sin}[2\pi/5]),\\
        \hat{\mathbf{r}}_2(n) &= (-1, \tau_n^{-1})\cdot(\hat{\mathbf{r}}_0,\hat{\mathbf{r}}_1),\\
        \hat{\mathbf{r}}_3(n) &= -(\tau_n^{-1}, \tau_n^{-1})\cdot(\hat{\mathbf{r}}_0,\hat{\mathbf{r}}_1),\\
        \hat{\mathbf{r}}_4(n) &= (\tau_n^{-1},-1)\cdot(\hat{\mathbf{r}}_0,\hat{\mathbf{r}}_1),\\
    \end{split}
\end{equation}
where $\tau_n = F_{n+1}/F_n$, $F_n$ is the $n$th Fibonacci number, and each grid is displaced by some phase $\gamma_i$ from the origin. 
The grids partition space into open regions which can be labeled uniquely by the five integers $\textbf{J} \equiv (j_0,j_1,\ldots, j_4)$.
Each point $\textbf{x}$ in some open region will lie between the lines $j_i$ and $j_i +1$ of the $i$th grid.
These open regions $\textbf{J}$ are then mapped to the vertices $\textbf{t}$ of a tiling by the transformation $\textbf{t} = \sum_{i=0}^4 j_i\hat{\textbf{r}}_i$.
This method yields a Penrose tiling if the sum of $\gamma_i$ is an integer multiple of the spacing between the parallel lines.
The number of tiles in the Penrose periodic approximants is determined by the index $n$ and different renditions of the same size of tiling can be obtained by changing the individual values of $\gamma_i$.
Here, we use 50 renditions of Penrose tiling periodic approximants with $n \in [13,17]$ ($N$ between 1149851 and 54018521).

In order to probe the hyperuniformity of the Fibonacci chains, period-doubling chains, and Penrose tilings described above using the excess spreadability, we map the one- and two-dimensional point configurations into packings.
Packings can be viewed as two-phase media, where phase $V_1$ is the void (pore) space between the particles, and the particle phase $V_2$ is the space occupied by the particles \cite{TorquatoBook}. 
To map chains (tilings) into two-phase media, the vertices of each link (tile) are decorated by identical rods (disks), of radius $a$ centered at the vertices, where $a$ is chosen such that the rods (disks) do not overlap.
The packing fraction of phase 2, $\phi_2$, is given as $\phi_2 = \rho v_1(a)$, where $v_1(a)$ is the volume of a rod or disk of radius $a$ defined in Eq. (\ref{volume}), and $\rho=N/V$ is the number density, where $N$ is the number of disks and $V$ is the total volume of the unit cell of the finite system.
For the 1D and 2D systems we use packing fractions $\phi_2$ = 0.35 and 0.25, respectively.
Note that $\alpha$ does not depend on the value of $a$ or $\phi_2$, so long as the rods (or disks) do not overlap.
Moreover, the packing fractions can take any value in the range $0 < \phi_2 \leq \phi_{2,max}$ where $\phi_{2,max}$ is the largest packing fraction for identical rods in $d=1$ or {identical} disks in $d=2$ {subject to the nonoverlap constraint}.
In particular, $\phi_{2,max}=4-2\tau\approx0.764$ for the 1D Fibonacci-chain packings and 
\begin{equation}
\begin{split}
    \phi_{2,max}&=\frac{\pi}{(2\tau)^2}\left(\frac{2}{\tau^2\sqrt{4-\tau^2}}+\left(\left(1-\frac{1}{\tau}
    \right)\frac{2\tau}{\sqrt{4-\frac{1}{\tau^2}}}\right) \right)\\
    & = \frac{2\pi}{\sqrt{130+58\sqrt{5}}}\approx0.390
\end{split}
\end{equation}
for 2D Penrose-tiling packings.
One can easily obtain the spectral density corresponding to any other packing fraction in the range above by simply multiplying our numerically computed spectral densities by the ratio $\phi_{2,new}/\phi_{2,i}$, where $\phi_{2,i}$ is the value used in this work.

\subsection{Calculation and extraction of $\alpha$ from the excess spreadability}\label{Fit_Meth}

To compute the spreadability, we first compute the structure factor using Eq. (\ref{scatinte}), where points of the lattice, $\textbf{r}$, are the points at the end of each link in one dimension or the vertices of the rhombi in the Penrose tiling in two dimensions.
The structure factor is then substituted into Eq. (\ref{spectral_nonoverlap}) to compute the spectral density, which is subsequently substituted into Eq. (\ref{spreadfromspectral}) to compute the excess spreadability. In the 2D Penrose case, we first perform a binned, angular average of the structure factor
to yield a radial spectral density via Eq. (\ref{spectral_nonoverlap}).

As shown in Eq. (\ref{longtime}), one can use the long-time behaviour of the excess spreadability to determine the value of the exponent $\alpha$.
To extract $\alpha$, we fit the long-time regime of the excess spreadability on a log-log scale to $y = c_0 - \frac{(d+\alpha)}{2} t$ over a period of time that minimizes the error on the measurement of $\alpha$.
To avoid finding a local error minimum that corresponds to a fit over a very small time period, we choose fit end points that are at least one decade apart in $t$.
We expect this fitting scheme to outperform the previous method \cite{haina} --- in which the point in the excess spreadability where the long-time scaling sets in is found via an iterative scheme and \textit{all} subsequent time points are fit to a set of trial functions --- especially for numerically sampled or experimental spreadabilities, because the new scheme removes the finite-size effects that occur at long times by construction, while the previous method requires one to do so manually.
To then determine $\alpha$ for the Fibonacci, period doubling, and Penrose media in the thermodynamic limit, we fit the $\alpha$ values extracted using the above method as a function of $1/N$ to a function of the form $\alpha = \hat{\alpha}-\frac{A}{N^B}$, where $\hat{\alpha}$, $A$, and $B$ are all free parameters, and $\hat{\alpha}$ symbolizes the value of $\alpha$ in the thermodynamic limit.
We find that this method can reproduce the theoretically known results for the Fibonacci and period doubling chains to within 0.02\% error compared to exact analytic results.

\section{Exact expressions for representative excess spreadability cases}\label{general}
Here, we review the spreadability results from Ref. \cite{TorquatoDiff} for the stealthy ($\alpha = \infty$) integer lattice, nonhyperuniform ($\alpha =0$) Debye random media, and a disordered hyperuniform medium ($\alpha = 2$). 
We later compare these results to the excess spreadabilities of the quasicrystaline and limit-periodic two-phase media (see Sec. \ref{FC}).

On long timescales, the spreadability distinguishes between the different systems according to their degree of order on large length scales. Torquato \cite{TorquatoDiff} showed that the spreadability of the nonhyperuniform Debye media scales as
\begin{equation}
    \mathcal{S}(\infty)-\mathcal{S}(t) \sim \frac{(d-1)!d\omega_d \phi_2}{(4\pi Dt/a^2 )^{d/2}}  - \frac{(d+1)!d\omega_d \phi_2}{(4\pi Dt/a^2 )^{(d+2)/2}},
\end{equation}
which implies asymptotic scaling of $\sim t^{-1/2}$ (which, according to Eq. (\ref{longtime}), corresponds to  $\alpha = 0$) for $d =1$. For the one-dimensional disordered hyperuniform medium, Torquato \cite{TorquatoDiff} found the scaling
\begin{equation}
    \mathcal{S}(\infty)-\mathcal{S}(t) \sim \frac{\phi_1}{4\pi^{1/2}( Dt/a)^{3/2}},
\end{equation}
which suggests asymptotic scaling of the form $\sim t^{3/2} \quad(\alpha = 2).$
For any periodic packing, Torquato showed that the spreadability is given exactly by \cite{TorquatoDiff}:
\begin{equation}
    \mathcal{S}(\infty)-\mathcal{S}(t) = \phi_2 \sum_{n=1} \mathcal{Z}(Q_n)  \frac{\Tilde{\alpha}_2(Q_n a)}{v_1(a)}e^{-Q^2_nDt}.
\end{equation}
Here $\mathcal{Z}(Q_n)$ is the expected coordination number at radial distance $Q_n$ and $a$ is the radius of the particles. 
Thus, at long times the excess spreadability of the integer lattice takes the form:
\begin{equation}
    \mathcal{S}(\infty)-\mathcal{S}(t) \sim e^{-Q^2_1Dt} \quad (Dt/a^2 \gg1),
\end{equation}
where $Q_1$ is the smallest positive Bragg wave number, which corresponds to $\alpha \rightarrow \infty$.

We note that for all two-phase media considered in this section the spectral density approaches the origin continuously.
The structure factor of the stealthy integer lattice is identically zero for wavenumbers smaller than the first Bragg peak (excluding forward scattering).
(For this reason, a stealthy system, strictly, speaking,  is not one in which $\alpha$ tends to infinity.) Hence, Eq. ($\ref{spectral_nonoverlap}$) implies that for a packing of identical spheres whose centers are at the lattice points, the spectral density too will be 0 before the first Bragg peak.
For the disordered hyperuniform medium the spectral density is given by \cite{TorquatoDiff}
\begin{equation}
    \frac{\Tilde{\chi}_{_V}(k) }{\phi_1 \phi_2} = \frac{4(ka)^2 a}{(ka)^4 +4},
\end{equation}
implying that for $k \ll 1$, the spectral density approaches 0 continuously as $\sim k^2$. 
The one-dimensional nonhyperuniform medium possesses a spectral density which at small wavevectors scales as \cite{TorquatoDiff}
\begin{equation}
   \Tilde{\chi}_{_V}(k) = \phi_1 \phi_2 \frac{2a}{\omega_0}\left[ 1- (ka)^2 + O(ka)^4 \right].
\end{equation}

\section{Hyperuniformity of the Period Doubling Chain}\label{PDC}

The limit-periodic period doubling chain, which also has scattering information consisting of a dense set of Bragg peaks, and a known hyperuniformity scaling exponent $\alpha=1$ \cite{primes}, is an ideal system for benchmarking the methods described above.
Here, we {compute the spectral density and subsequently the excess spreadability for} the two-phase media derived from period doubling chains generated using the substitution method (described in Sec. \ref{Methods}) with 13 sizes between $N = 21845$ and 22369621 to show, for the first time, that the excess spreadability can be used to accurately extract $\alpha$ for the period doubling chain.
In the $N\rightarrow\infty$ limit, the tiling is a union of periodic systems, which is termed limit-periodic.
In this limit, the structure factor associated with the $S$ links is given by\cite{primes,baake2011diffraction}:
\begin{equation}
\begin{aligned}\label{lim-per-sk}
    S(\text{k}) = \frac{4\pi}{3}  &\sum_{m=1}^{\infty} \delta(\text{k} - 2\pi m) + \\
    \frac{4\pi}{3}&\sum_{n =1}^{\infty} 2^{-2n} \sum_{m=1}^{\infty} \delta \left( k-\frac{(2m-1)\pi}{2^{n-1}}  \right), 
\end{aligned}
\end{equation}
assuming unit lattice spacing, which shares properties with $S(k)$ for quasiperiodic systems, most importantly the presence of dense Bragg peaks. 
In Ref. \citenum{primes}, Torquato et al. showed that Eq. (\ref{lim-per-sk}) scales linearly in the vicinity of the origin, i.e., $\alpha = 1$, and thus the period-doubling chain is class II hyperuniform.
As stated in Sec. \ref{Background} B, $\Tilde{\chi}_{_V}(k)$ will inherit $\alpha = 1$ from the $S(k)$ given in Eq. (\ref{lim-per-sk}).
Thus, we expect our fitting scheme to extract $\alpha = 1$ from the long-time scaling of the excess spreadability.

\begin{figure}
    \centering
    \includegraphics[width=0.45\textwidth]{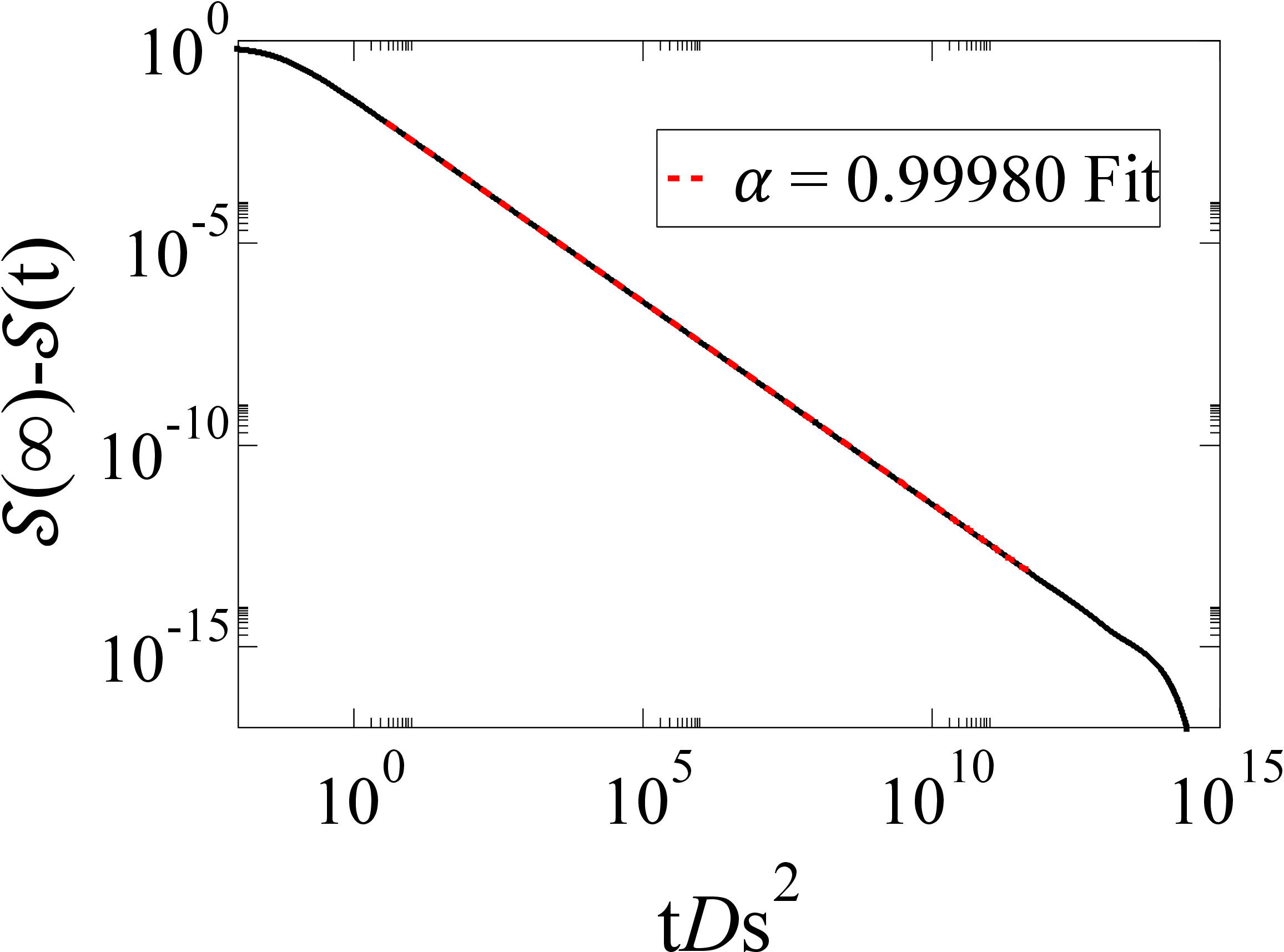}
    \caption{An example excess spreadability $\mathcal{S}(\infty)-\mathcal{S}(t)$ curve as a function of dimensionless time $tDs^2$, scaled by the diffusion coefficient $D$ and specific surface $s$, on a log-log scale for the two-phase medium derived from the period-doubling chain with a packing fraction $\phi_2=0.35$ and $N = 22369621$. The dashed red (gray) line shows the error-minimizing fit with $\alpha = 0.99980$.}
    \label{PD_SpreadEx}
\end{figure}

\begin{figure}
    \centering
    \includegraphics[width=0.45\textwidth]{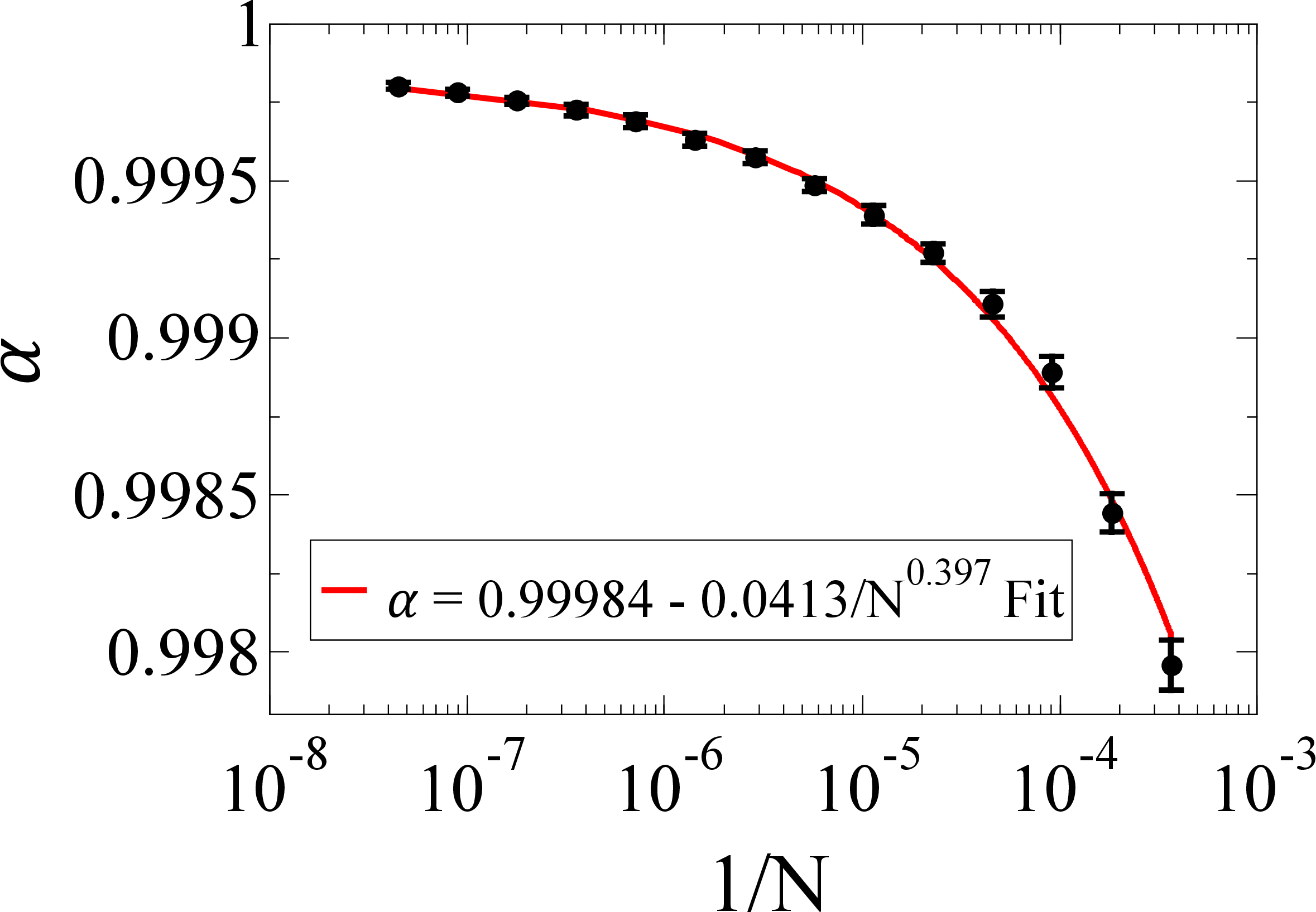}
    \caption{A semi-log plot of the values of $\alpha$ extracted from the excess spreadability as a function of $1/N$, where $N$ is the number of particles in the media derived from the period doubling chains. The solid red (gray) line shows the fit to $\alpha = \hat{\alpha}-\frac{A}{N^B}$, where $\hat{\alpha} = 0.99984\pm 0.00001$ is the value of $\alpha$ in the thermodynamic limit.}
    \label{PD_FSS}
\end{figure}

Figure \ref{PD_SpreadEx} shows the excess spreadability curve for the largest of the period doubling chain systems considered in this work, which clearly exhibits a large range over which the power law scaling corresponds to a hyperuniformity scaling exponent $\alpha = 0.99980\pm 0.00001 1$.
The exponentially fast drop-off at very large $t$ is a finite-size effect associated with lack of scattering information between $k = 0$ and $k = k_{min}$, where $k_{min} = 2\pi / L$ is the smallest admissible wavenumber, and $L$ is the length of the medium.
Clearly, the error-minimizing fit scheme described in Sec. \ref{Methods} is insensitive to these finite-size effects.
Figure \ref{PD_FSS} shows the values of $\alpha$ extracted from the excess spreadability curves of the 13 period doubling chain systems as a function of $1/N$.
The error bars on the individual $\alpha$ measurements are those extracted from the error-minimizing fits.
Fitting this data to $\alpha = \hat{\alpha}-\frac{A}{N^B}$ reveals that, in the thermodynamic limit, $\alpha$ approaches a value of $\hat{\alpha}=0.99984\pm
0.00001 $, which is within $0.02\%$ of the theoretically known value of $\alpha = 1$, and whose error comes from the standard deviation of the $\hat{\alpha}$ estimate.
Thus, we have confirmed that the proposed fitting method is able to accurately extract the expected value of $\alpha$ from the excess spreadability.

\section{Hyperuniformity of the Fibonacci Chain}\label{FC}

Having established that the proposed methodology can accurately extract $\alpha$ from the spreadability of a medium whose $S(k)$ is discontinuous in the vicinity of the origin, we will show that it also works for quasicrystalline media.
{We compute the spectral density and subsequently the excess spreadability} for packings derived from 15 Fibonacci chains with $N$ between 2584 and 514229 particles.
O\u{g}uz et al. \cite{quasihyperuniformity} have analytically shown that $S(k)$ for the Fibonacci chain scales with $\alpha = 3$ in the vicinity of the origin, meaning it is class I hyperuniform.

\begin{figure}
    \centering
    \includegraphics[width=0.45\textwidth]{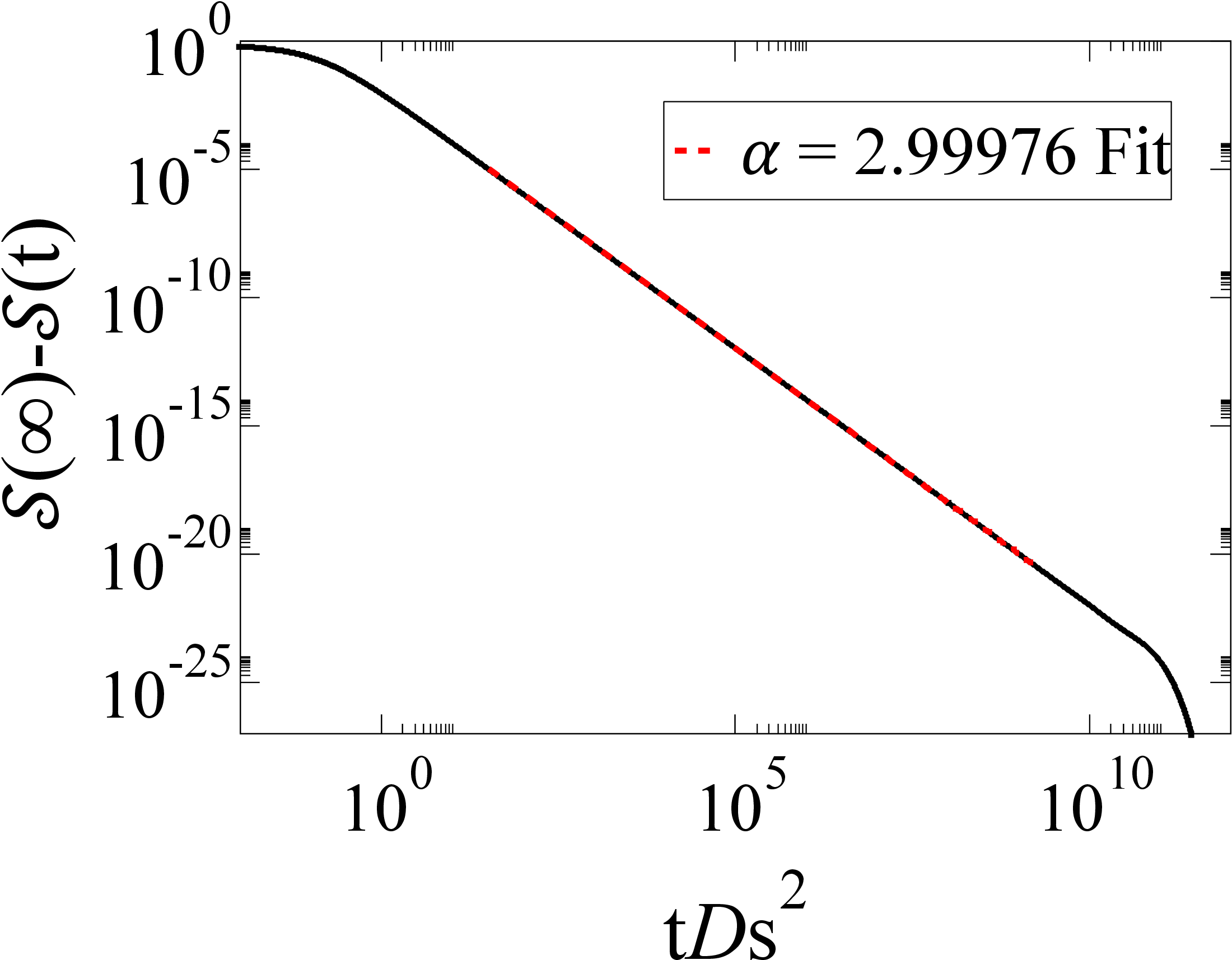}
    \caption{An example excess spreadability $\mathcal{S}(\infty)-\mathcal{S}(t)$ curve as a function of dimensionless time $tDs^2$, scaled by the diffusion coefficient $D$ and specific surface $s$, on a log-log scale for the two-phase medium derived from the Fibonacci chain with a packing fraction $\phi_2=0.35$ and $N = 514229$. The dashed red (gray) line shows the error-minimizing fit with $\alpha = 2.99976$.}
    \label{FC_SpreadEx}
\end{figure}

\begin{figure}
    \centering
    \includegraphics[width=0.45\textwidth]{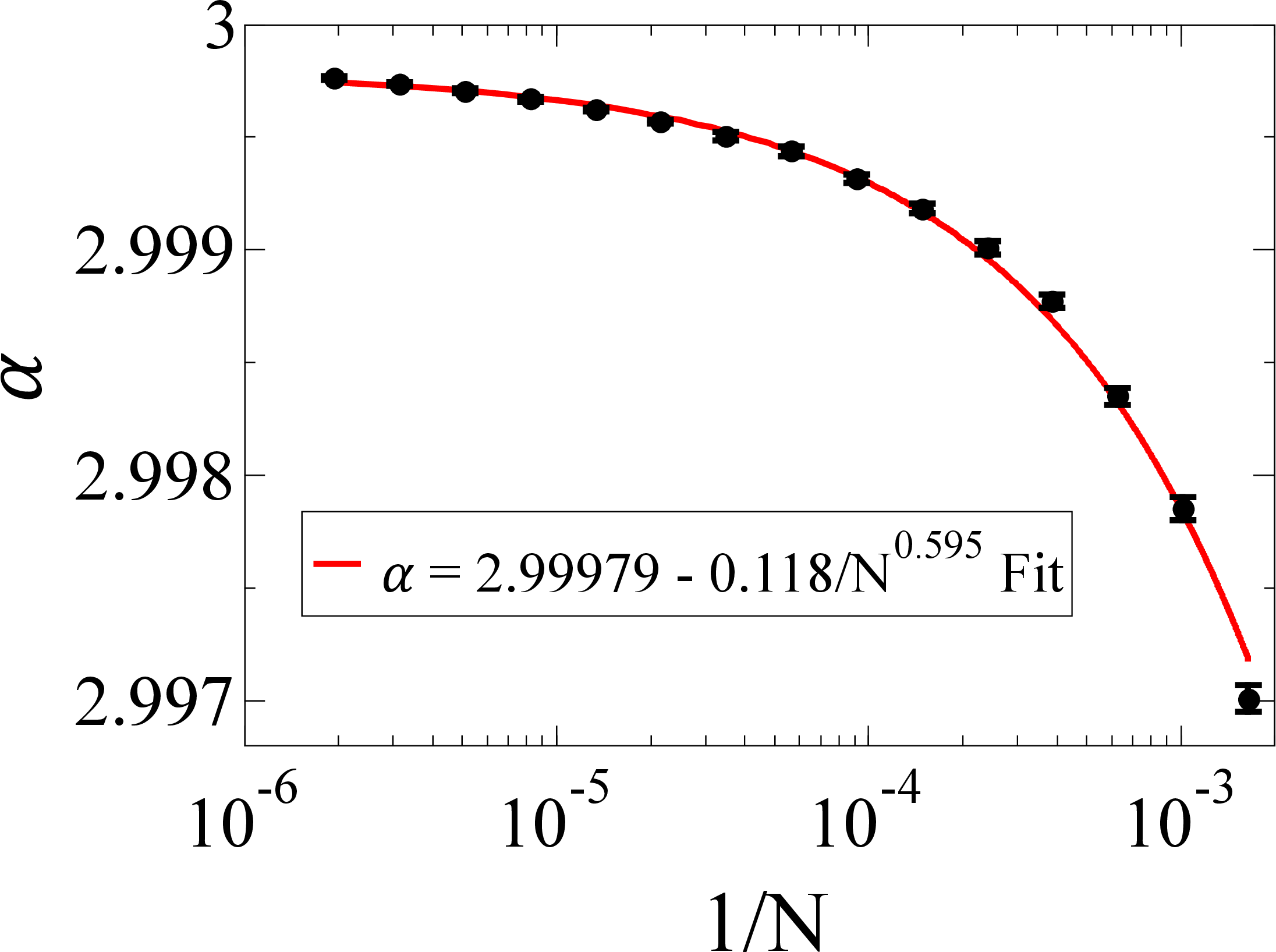}
    \caption{A semi-log plot of the values of $\alpha$ extracted from the excess spreadability as a function of $1/N$, where $N$ is the number of particles in the media derived from the period doubling chains. The solid red (gray) line shows the fit to $\alpha = \hat{\alpha}-\frac{A}{N^B}$, where $\hat{\alpha} = 2.99979\pm 0.000008$ is the value of $\alpha$ in the thermodynamic limit.}
    \label{FC_FSS}
\end{figure}

Figure \ref{FC_SpreadEx} shows the excess spreadability curve for the largest of the Fibonacci chain systems considered in this work, which has a large range of times over which the power law scaling corresponds to $\alpha = 2.99976 \pm 0.000008$.
One can also observe the same exponential drop-off at large times as the period doubling chain excess spreadability curve, due to the same finite size effect.
Figure \ref{FC_FSS} shows the values of $\alpha$ extracted from the excess spreadabilities for all of the Fibonacci chains as a function of $1/N$.
The error bars on the individual $\alpha$ measurements are those extracted from the error-minimizing fits.
Fitting this data to $\alpha = \hat{\alpha}-\frac{A}{N^B}$ reveals that, in the thermodynamic limit, $\alpha$ approaches a value of $\hat{\alpha}=2.99979\pm
0.000008$, which is within $0.01\%$ of the theoretically known value of $\alpha = 3$, and whose error comes from the standard deviation of the $\hat{\alpha}$ estimate.
This result confirms the ability of our $\alpha$ extraction method to accurately extract $\alpha$ for quasiperiodic two-phase media whose spectral functions are dense and discontinuous in the vicinity of the origin.
\begin{figure}
    \centering
    \includegraphics[width=0.45\textwidth]{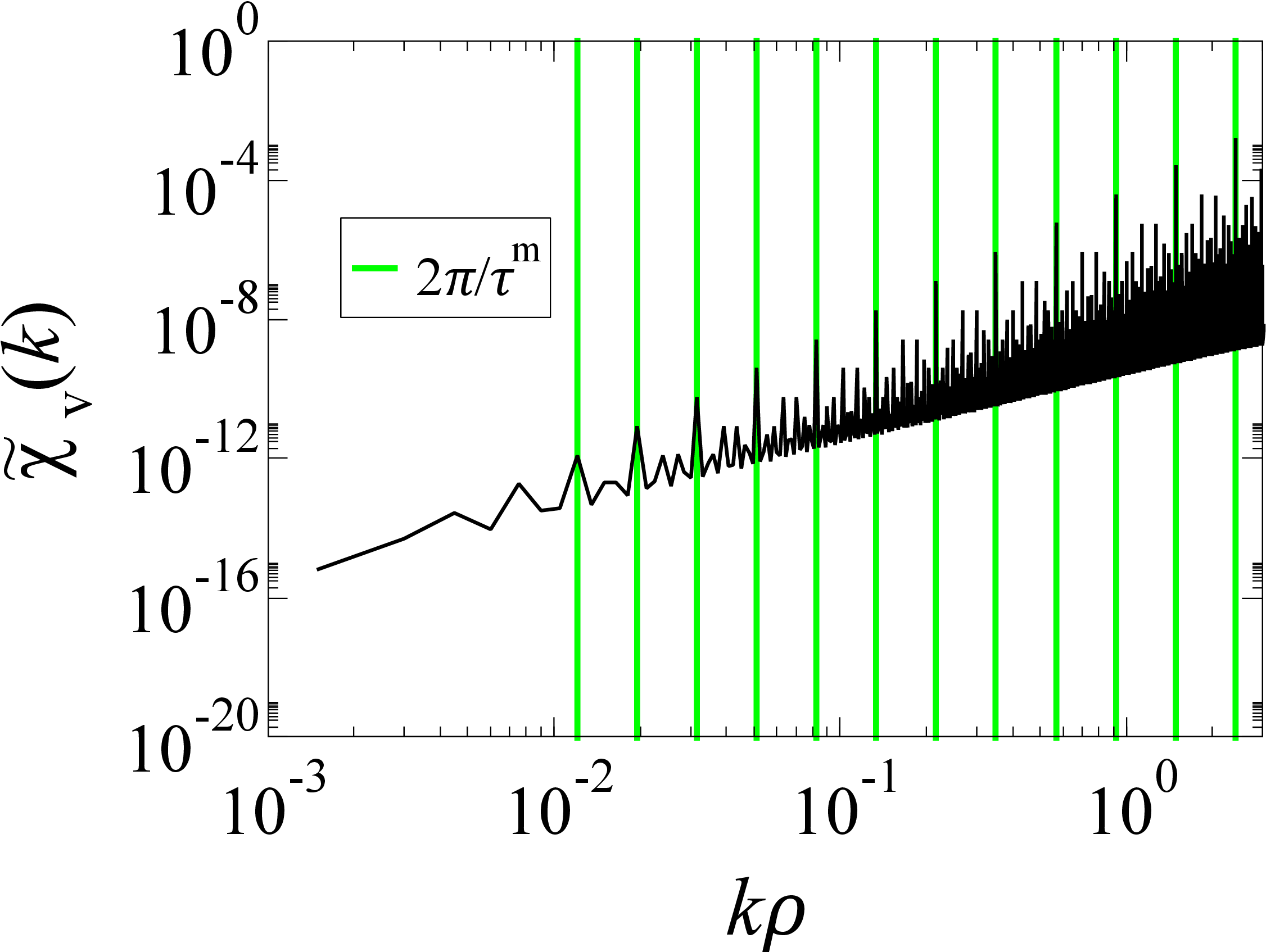}
    \caption{A log-log plot of the spectral density $\Tilde{\chi}_{_V}(k)$ as a function of the wavenumber $k$ scaled by the number density $\rho$ of a two-phase medium derived from a Fibonacci chain with a packing fraction $\phi_2 = 0.35$ and $N = 4181$. The vertical green (gray) lines denote the wavenumbers $k = 2\pi/\tau^m$, where $\tau = \frac{1 + \sqrt{5}}{2}$, and $m\in [2,13]$ is an integer. Note that the vertical green (gray) lines coincide with the tallest peaks relative to those surrounding them.}
    \label{Fib_cut}
\end{figure}

We note here that the substitution tilings considered here are self-similar.
This self-similarity manifests in the small-$k$ behavior of $S(k)$ and $\Tilde{\chi}_{_V}(k)$, where one can observe patterns in the peak heights that repeat with a regular period when $k$ is scaled logarithmically.
In the thermodynamic limit, this pattern will repeat infinitely many times as $k\rightarrow0$.
We will now show that, due to this self-similarity, it is possible to truncate the small-$k$ region of the spectral density at a wavenumber $k_{cutoff}$ and recover a value of $\alpha$ from the corresponding excess spreadabilty equal to that of the untruncated spectral density within a small error that decreases as $N$ increases.
Figure \ref{Fib_cut}, which shows $\Tilde{\chi}_{_V}(k)$ for a two-phase medium derived from an $N = 4181$ Fibonacci chain, demonstrates that this periodic pattern of peak heights occurs at integer powers of $1/\tau$ (modulo a factor of $2\pi$) for Fibonacci chains.
For this particular chain, when considering the entire small-$k$ region of the spectral density, fitting the corresponding excess spreadability yields $\alpha = 2.99901\pm   0.00003$.
As an example, we find that by considering only $k > k_{cutoff}=2\pi/\tau^{12}$, which corresponds to only considering real-space length scales that are smaller than 1/10 of the entire system size, we can extract $\alpha = 2.99971 \pm 0.000008$, which differs from the untruncated value by less than 0.2\%.
The difference between the truncated and untruncated $\alpha$ values diminishes as the system size increases.
It is important to note that this is only possible with structures whose structure factors are self-similar at small $k$, in particular, one should not expect this procedure to yield the correct $\alpha$ for general disordered hyperuniform media.

\begin{figure}[b]
\centering
\includegraphics[width=0.9\linewidth]{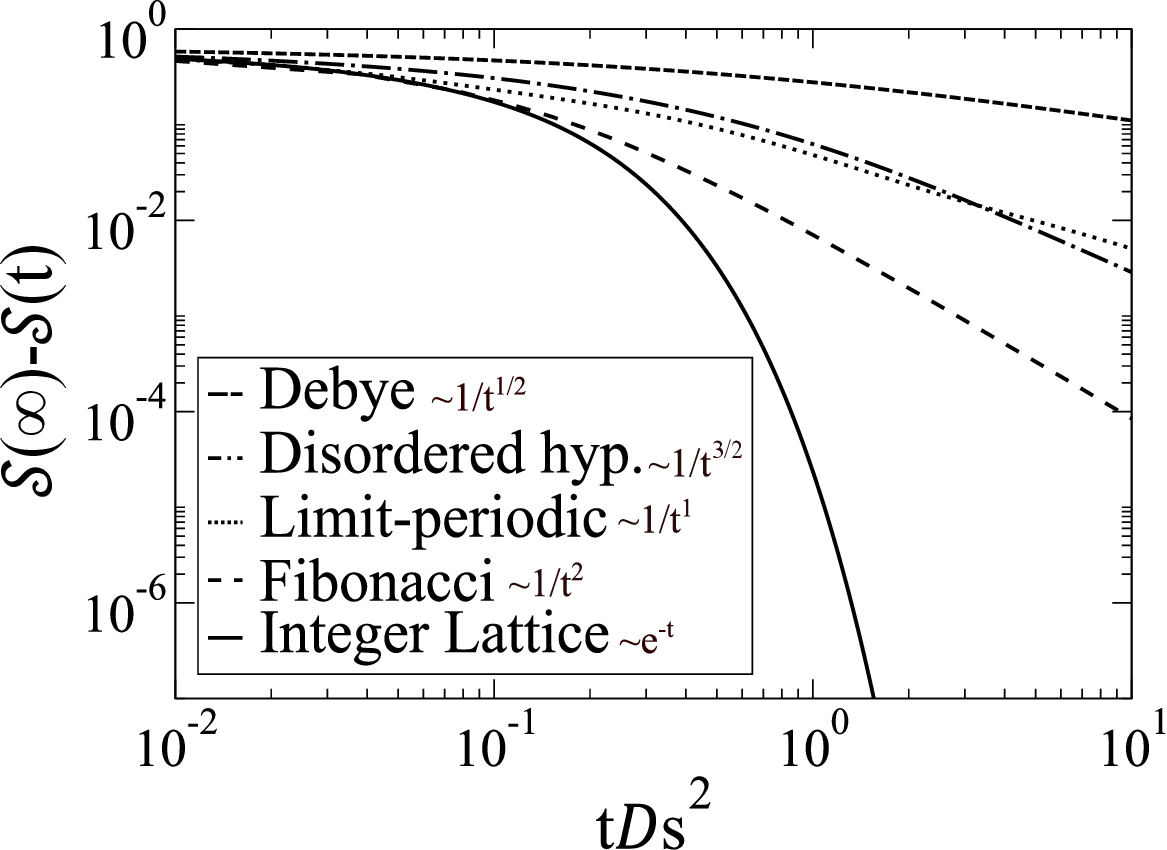}
\caption{\small Log-log plot of the excess spreadability, $\mathcal{S}(t)-\mathcal{S}(\infty)$,  for dimensionless time $10^{-2}\le$t$D$s$^2$$\le10^1$ of different 1D systems with $\phi_2 = 0.35$. The long-time behavior of the spreadability is determined by the degree of large-scale order of the systems where the fastest decay ($\sim e^{-t}$) is for the integer lattice, which is the most ordered system, and the slowest ($\sim t^{-1/2}$) is for the Debye, which is the most disordered at large length scales. The rest of the systems fall between the two. Debye, disordered hyperuniform, and integer lattice taken from \cite{TorquatoDiff}. \label{different_1D_systems}}
\end{figure}

Figure \ref{different_1D_systems} compares the excess spreadabilities of the Fibonacci and period-doubling chains to those of the 1D two-phase media described in Sec. \ref{general}. 
Equation (\ref{longtime}) implies long-time scaling of $t^{-2}$ and $t^{-1}$ for the limit-periodic and Fibonacci chains with $\alpha= 1$ and $\alpha = 3$, respectively.
In Fig. \ref{different_1D_systems}, one can clearly see how the different values of $\alpha$ translate to different decay behavior in the large-time regime of the excess spreadability.
The fastest decay of the spreadability ($\sim e^{-t}$) occurs for a two-phase medium derived from the integer lattice, which is stealthy with $\alpha = \infty$.
The slowest decay ($\sim t^{-{1/2}}$) occurs for the nonhyperuniform Debye random medium, which is the most disordered system examined here at large length scales, with $\alpha = 0$. Falling between these two, the Fibonacci decays with $\sim t^{-2}$ ($\alpha = 3$), the disordered hyperuniform medium with $\sim t^{-3/2}$ ($\alpha = 2$), and the period-doubling with $\sim t^{-1}$ ($\alpha = 1$).

The agreement between the theoretical values of $\alpha$ for the two-phase media derived from the Fibonacci and period-doubling chains and the results obtained here using the procedure given in Sec. \ref{Methods} suggest that it can accurately extract $\alpha$ from two-phase media derived from quasicrystals and limit-periodic point patterns.
Moreover, Fig. \ref{different_1D_systems} clearly demonstrates that the spreadability can be used to accurately extract the value of $\alpha$, for materials of any hyperuniformity class, including nonhyperuniform materials and materials whose $S(k)$ is discontinuous in the vicinity of the origin.
We note that while $Z(k)$ has been used in the past to characterize the hyperuniformity of the Fibonacci chain \cite{quasihyperuniformity}, it is difficult to do so for arbitrary systems with dense Bragg peaks in the vicinity of the origin because one requires the constants derived from an analytical $S(k)$ (described in Sec. \ref{Intro}) to fit $Z(k)$ reliably.

{In addition to examining the small-$k$ behavior of the 1D Fibonacci-chain packing, like we have done above, one can consider the large-$k$, i.e., small-wavelength, behavior of the Fibonacci-chain packing.}
{In Appendix A, we present the numerically computed spectral density of a Fibonacci-chain packing for a wide range of $k$ values, and show that the large-$k$ scaling behavior is controlled by the power law scaling given in Eq. (17).}

\begin{figure}
    \centering
    \includegraphics[width=0.45\textwidth]{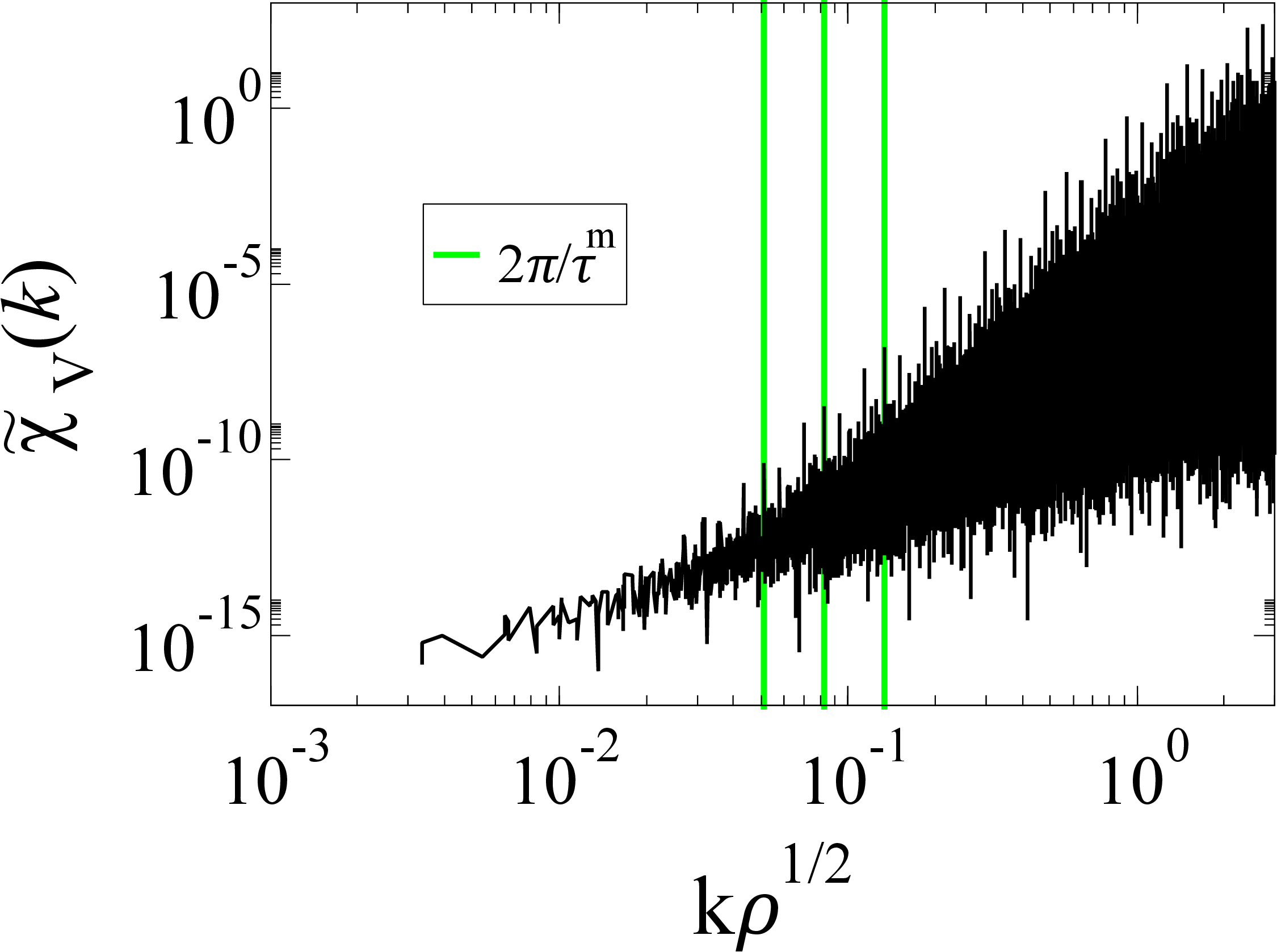}
    \caption{A log-log plot of the spectral density $\Tilde{\chi}_{_V}(k)$ as a function of the wavenumber $k$ scaled by the number density $\rho$ of a two-phase medium derived from a Penrose tiling periodic approximant chain with a packing fraction $\phi_2 = 0.25$ and $n = 14\;(N = 3010349)$. The vertical green (gray) lines denote the wavenumbers $k = 2\pi/\tau^m$, where $\tau = \frac{1 + \sqrt{5}}{2}$, and $m\in [8,10]$ is an integer. Note that the vertical green (gray) lines coincide with the tallest peaks relative to those surrounding them.}
    \label{Pen_cut}
\end{figure}
\section{Hyperuniformity of the Penrose Tiling}\label{PT}

Having established that the procedure described in Sec. \ref{Methods} can extract $\alpha$ from 1D systems whose structure factor consists of a dense set of Bragg peaks, we now use the excess spreadability to extract $\alpha$ for the 2D Penrose tiling. 
Here, we {compute the spectral densities and subsequently the excess spreadabilties of} 50 renditions of packings derived from Penrose tiling periodic approximants with $n\in[13,17]$, i.e., $N$ between 1149851 and 54018521 (see Sec. \ref{Methods} for more details).

Penrose tilings are also self-similar \cite{penrose1974role}, so one would expect there to be a repeating pattern of peaks in the small-$k$ region as $k \rightarrow 0$, similar to those seen in Fig. \ref{Fib_cut}.
In practice, however, we find that there is a breakdown of this self-similarity in the spectral density at small $k$, i.e., at large length scales.
Figure \ref{Pen_cut} shows a transition from clear ``triplets'' of peaks at integer powers of $1/\tau$ (modulo a factor of $2\pi$) to a regime where there are no well-defined peak triplets that approaches the origin more slowly.
We attribute this breakdown in the self-similarity at large length scales to the finite size of our Penrose tiling periodic approximants.
Moreover, the length scale associated with this breakdown in self-similarity becomes a smaller fraction of the size of the periodic approximant as the approximant number $n$ increases.
Given that the value of $\alpha$ extracted from the excess spreadability is not altered significantly by imposing a $k_{cutoff}$ in the spectral density for the self-similar Fibonacci chains, we do so here in order to remove this finite size effect.
For the system sizes examined here, we find that $k_{cutoff} = 2\pi / \tau^{n-4}$ is effective in mitigating these finite size effects.

\begin{figure}
    \centering
    \includegraphics[width=0.45\textwidth]{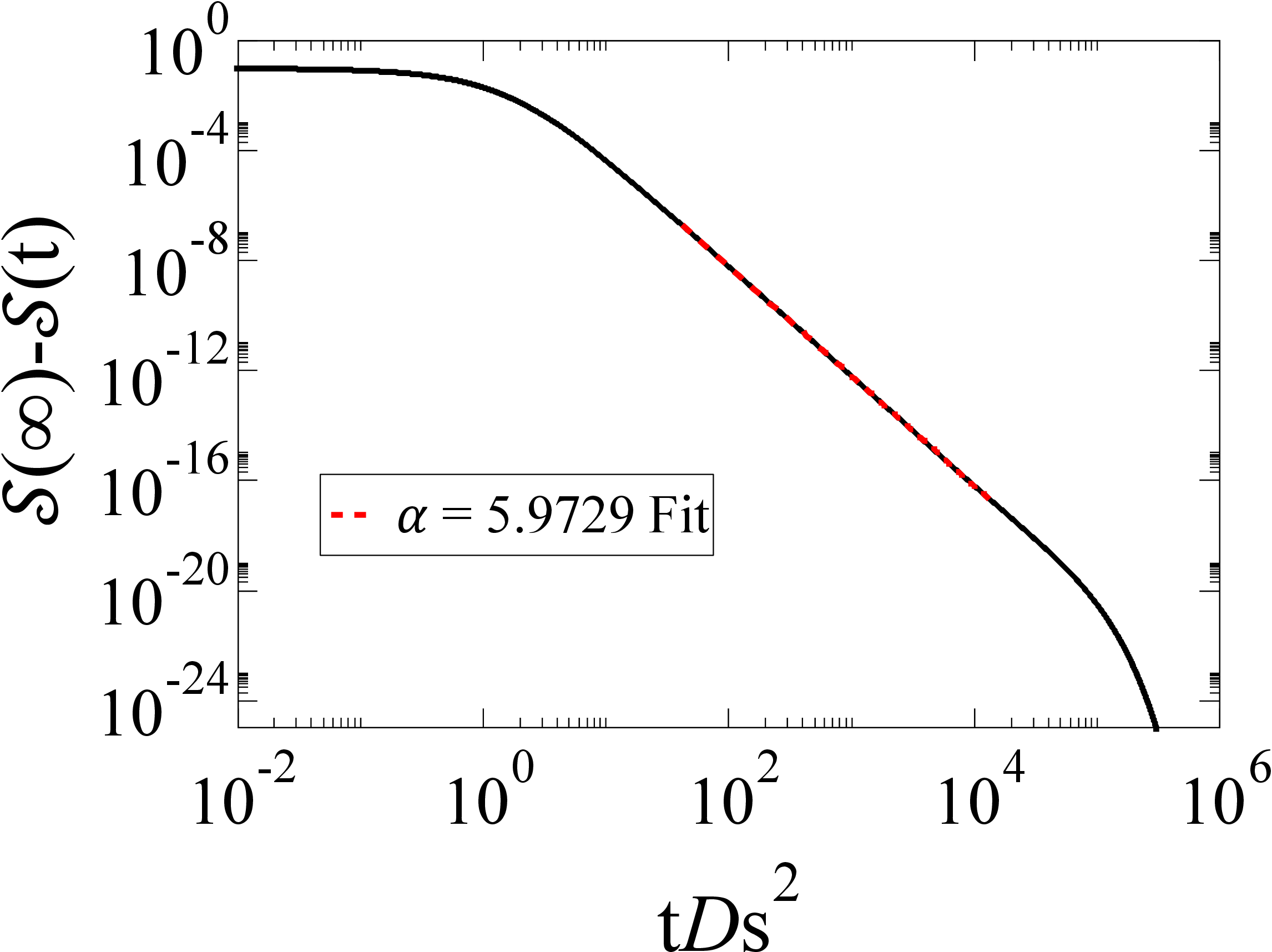}
    \caption{An example excess spreadability $\mathcal{S}(\infty)-\mathcal{S}(t)$ curve as a function of dimensionless time $tDs^2$, scaled by the diffusion coefficient $D$ and specific surface $s$, on a log-log scale for the two-phase medium derived from Penrose tiling periodic approxiamnt with a packing fraction $\phi_2=0.25$ and $n = 17\;(N = 54018521)$. The dashed red (gray) line shows the error-minimizing fit with $\alpha = 5.9729$.}
    \label{PT_SpreadEx}
\end{figure}
\begin{figure}
    \centering
    \includegraphics[width=0.45\textwidth]{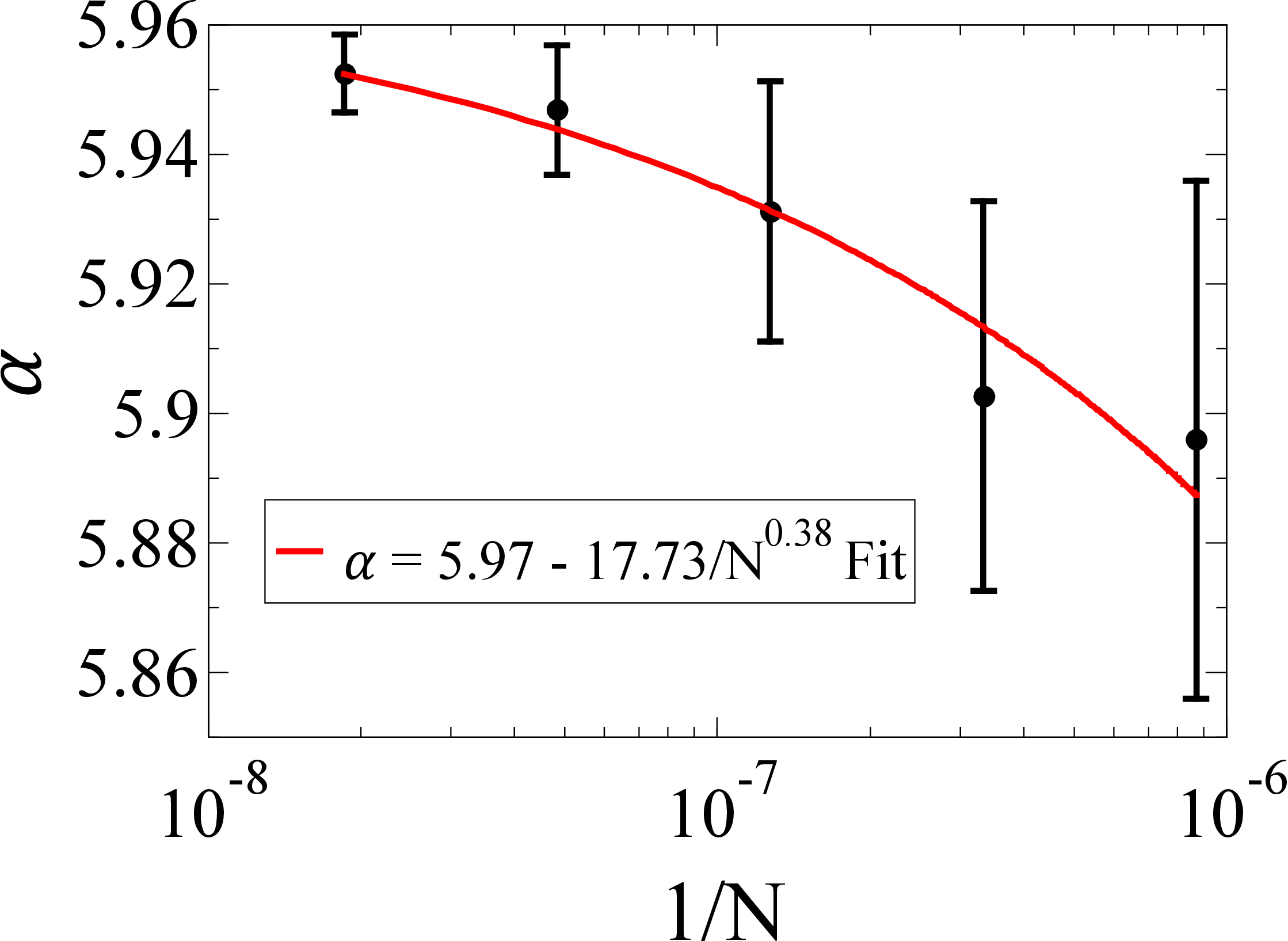}
    \caption{A semi-log plot of the values of $\alpha$ extracted from the excess spreadability as a function of $1/N$, where $N$ is the number of particles in the media derived from the Penrose tilings. The solid red (gray) line shows the fit to $\alpha = \hat{\alpha}-\frac{A}{N^B}$, where $\hat{\alpha} = 5.97\pm 0.06$ is the value of $\alpha$ in the thermodynamic limit.}
    \label{PT_FSS}
\end{figure}

Figure \ref{PT_SpreadEx} shows an example excess spreadability curve for a rendition of an $n=17$ Penrose tiling periodic approximant.
As in the 1D cases, there is a large region over which $\alpha$ remains stable at a constant value, specifically $\alpha = 5.9729\pm0.0003$, followed by an exponential drop-off.
Note also that imposing $k_{cutoff}$ has mitigated the effect of the non-self-similar regime of the spectral density on the excess spreadability, i.e., there is only a single power-law scaling regime visible in Fig. \ref{PT_SpreadEx}, which corresponds to the self-similar portion of the spectral density.
Figure \ref{PT_FSS} shows the ensemble-averaged values of $\alpha$ extracted from each of the 5 ensembles of Penrose tiling periodic approximants as a function of $1/N$.
The error bars on the $\alpha$ values come from the standard deviation of the individual $\alpha$ values extracted from all renditions of a particular $N$, which is significantly larger than the error on any specific measurement of $\alpha$.
Fitting this data to $\alpha = \hat{\alpha}-\frac{A}{N^B}$ shows that $\alpha$ approaches a value of 5.97$\pm0.06$ in the thermodynamic limit, whose error comes from the standard deviation of the $\hat{\alpha}$ estimate. The validation our numerical procedure to extract the expected exact values
of $\alpha$ for the 1D self-similar packings described in Secs. \ref{PDC} and \ref{FC} as well as 
the highly plausible expectation that the Penrose tiling and other Class I tilings obtained 
by standard projection from higher dimensional periodic lattices have integer values of $\alpha$, strongly supports the conjecture
 $\alpha \to 6$ in the thermodynamic (infinite-volume) limit.
 
{Following what we have done for the 1D Fibonacci-chain packings, we can additionally consider the behavior of the large-$k$, i.e., small-wavelength behavior of the Penrose-tiling packing.}
{In Appendix A, we present the numerically computed spectral density of a Penrose-tiling packing for a wide range of $k$ values, and show that the large-$k$ scaling behavior is controlled by the power law scaling given in Eq. (17).}

\section{Conclusions and Discussion \label{Discussion}}
In this work, we used the spreadability to characterize the hyperuniformity and time dependent diffusion properties of one- and two-dimensional two-phase media whose spectral densities comprise dense sets of Bragg peaks.
In particular, we produced sphere packings where the sphere centroids were the vertices of the Fibonacci quasicrystal and period-doubling limit-periodic 
chain in one dimension and the Penrose quasicrystal in two dimensions.
To extract $\alpha$ from the excess spreadabilities of these media with a dense set of Bragg peaks as $k\rightarrow0$, we formulated a new fitting scheme.
Specifically, we  showed that, by leveraging the self-similarity of 1D substitution tilings and the Penrose tiling, one can 
ignore a portion of the small-$k$ region of the spectral density and obtain a value of $\alpha$ from the excess spreadability that is equal to the $\alpha$ obtained from the excess spreadability computed using the complete spectral density,
with a small degree of error. We found that our procedure accurately extracts the theoretical values of $\alpha$ for the Fibonacci quasicrystal and period-doubling limit-periodic sequences to within 0.01\% and 0.02\% error, respectively.
We then used the excess spreadability to measure $\alpha$ for the Penrose quasicrystal, and found that, in the thermodynamic limit, $\alpha = 5.97 \pm 0.06$.
The accuracy of our numerical procedure and plausible theoretical arguments  noted in Sec. \ref{PT}, implies the exact result  $\alpha=6$ for the Penrose-tiling
packing and hence the Penrose quasicrystal point pattern. In future work, we will provide more examples and endeavor to construct   a rigorous proof
of this conjecture.

The accuracy of the $\alpha$ values extracted from the two-phase media considered here, whose spectral densities are discontinuous in the vicinity of the origin, further demonstrates the utility of the spreadability to extract the hyperuniformity scaling exponents for an extremely broad class of two-phase media including a larger class of quasicrystal-derived materials.
Specifically, we expect the methods described above to be able to easily extract a value of $\alpha$ for quasicrystalline and limit-periodic media in any space dimension that are self-similar, including, e.g., the prime numbers, which are effectively limit-periodic
\cite{primes}.
These structural characteristics affect a wide variety of physical properties of quasicrystalline and limit-periodic media including: the spreadability of diffusion information \cite{TorquatoDiff}, the nuclear magnetic resonance and magnetic resonance imaging measurements \cite{mag1, mag2, mag3}, rigorous upper bounds on the fluid permeability \cite{AWR}, the electromagnetic wave characteristics beyond the quasistatic regime \cite{qs1}, and photonic band gaps \cite{klatt2022wave}.

To date, $Z(k)$ has been used to  extract $\alpha$ from point patterns whose $S(\mathbf{k})$ are discontinuous with dense support for arbitrarily small $k$, which includes quasicrystalline and limit-periodic point patterns (see, e.g. Ref. \citenum{quasihyperuniformity}).
To extract $\alpha$ by fitting $Z(k)$ for the systems in Ref. \citenum{quasihyperuniformity}, coefficients derived from analytical expressions of $S(k)$ of the point patterns were used to guide the fit.
However, it is not known what these coefficients are for general quasicrystalline and limit-periodic point patterns, and directly fitting $Z(k)$ without them is problematic because $Z(k)$ oscillates as a function of $\log(k)$ as $k\rightarrow0$ for $S(k)$ with dense Bragg peaks \cite{quasihyperuniformity}.
The spreadability, as we have shown here, does not require an analytical expression for $S(k)$ or $\Tilde{\chi}_{_V}(k)$ to accurately extract a value of $\alpha$ from these same types of systems, which makes it more broadly applicable to other media whose spectral density is composed of dense Bragg peaks.
Lastly we note that while $Z(k)$ is essentially an arbitrary choice of smoothing of the structure factor, the excess spreadability is a physical, measurable, quantity that can be measured experimentally even when scattering information is not available \cite{SKOLNICK}.

To build on the characterization of the diffusion properties of the two-phase quasicrystalline media presented here, future work should focus on the characterization of their other physical properties e.g., the effective dynamic dielectric constant \cite{qs1,opt2,opt3,opt4}, dynamic effective elastic constant \cite{el1}, fluid permeability \cite{AWR}, and trapping constant \cite{AWR}, all of which can be estimated from the spectral densities.
Such a mapping of quasicrystalline point patterns to two-phase media could also allow for the characterization of the hyperuniformity of a wider class of 2D point patterns generated using the Generalized Dual Method \cite{lin2017hyperuniformity,dealgebraic,socolar1985quasicrystals}, as well as 3D quasicrystalline point patterns.
The characterization of the small- and intermediate-length scale properties of two-phase materials derived from substitution tilings and the Generalized Dual Method \cite{lin2017hyperuniformity,dealgebraic,socolar1985quasicrystals} using the excess spreadability is also of interest.

Finally, it is instructive to remark on the fundamental and practical implications of the small-wavenumber scaling of the spectral density for 1D Fibonacci-chain and  2D Penrose-tiling packings, given by ${\tilde \chi}_{_V}(k) \sim k^3$ and  ${\tilde \chi}_{_V}(k) \sim k^6$, respectively.
First, we note that as one goes from such quasiperiodic 1D to 2D packings, the hyperuniform systems become more ``stealthy-like", i.e., the spectral function gets flatter as $d$ increases ($\alpha$ goes from 3 to 6) \cite{To15,torquato2016disordered,steal_foot}, which has implications for attenuation of waves at long wavelengths \cite{qs1,opt3,opt4,el1}, for example. 
Importantly, the fact that $\alpha=3$ for 1D Fibonacci-chain packings means that ${\tilde \chi}(k)$ is nonanalytic at the origin, since it is an odd power, which in  turn implies
the decay of the corresponding direct-space autocovariance function $\chi_{_V}(r)$
to zero is controlled by  the inverse power-law $1/r^4$ \cite{hyperuniformstatesofmatter}.
Note that a spectral density that is analytic at the origin possesses a series expansion about $k=0$ that contains only even powers of $k$, which implies
that  $\chi_{_V}(r)$ decays to zero  exponentially fast or faster \cite{hyperuniformstatesofmatter}.
At first glance, our conjectured result for Penrose packings that $\alpha=6$, an even power in $k$,  seems to suggest that ${\tilde \chi}_{_V}(k)$ is analytic at the origin, but this conclusion would be inconsistent with the corresponding nonanalytic behavior for the 1D Fibonacci packings, since the latter patterns are encoded in the former. 
Moreover, a leading order even power does not ensure analyticity becuase higher order terms may include odd powers \cite{hyperuniformstatesofmatter, nonan_foot}.
We have verified that this situation is indeed the case by performing
numerical fits of the spectral density for small $k$ and find that
there is a substantial $k^7$ term, whose coefficent is of order one but about
an order of magnitude larger than that of the coefficient $k^6$.
This small-$k$ behavior means that $\chi_{_V}(r)$  decays to zero like $1/r^9$, which is much faster than the decay rate for 1D Fibonacci packings.
We note that 4D maximally random jammed hypersphere packings provides another example of a packing with an even-valued $\alpha$ and nonanalytic spectral density \cite{Maher_Hyperspheres}.

~

\begin{acknowledgments}
The authors thank A. Zentner, H. Wang, and M. Skolnick for insightful discussions. PJS thanks  C. Vafa and the High  Energy Physics group in the Department of Physics at Harvard University for graciously hosting him during his sabbatical leave. This research was sponsored by the Army Research Office and was accomplished under Cooperative Agreement No. W911NF-22-2-0103.
\end{acknowledgments}
~

\section*{Appendix A: Large-$k$ and Small-$k$ Scaling of the Spectral Densities of Quasicrystalline Packings}

Here we present our numerically determined spectral densities for the 1D Fibonacci-chain and 2D Penrose-tiling packings for a wide range of wavenumbers.
Figure 11 clearly shows that the large-$k$ regions of the spectral densities of the two-phase packings derived from the Fibonacci and Penrose quasicrystals decay like $k^{-2}$ and $k^{-3}$, respectively, as required by Eq. (17).
The presence of such scaling supports the accuracy of our numerical calculations.
In these plots, one can also clearly see that the spectral density of the 2D Penrose-tiling packing has a substantially faster decay in the small-$k$ region ($\alpha=6$) than that of the 1D Fibonacci-chain packing ($\alpha=3$).

\begin{figure}[ht]
    \centering
    \subfigure[]{\includegraphics[width=0.45\textwidth]{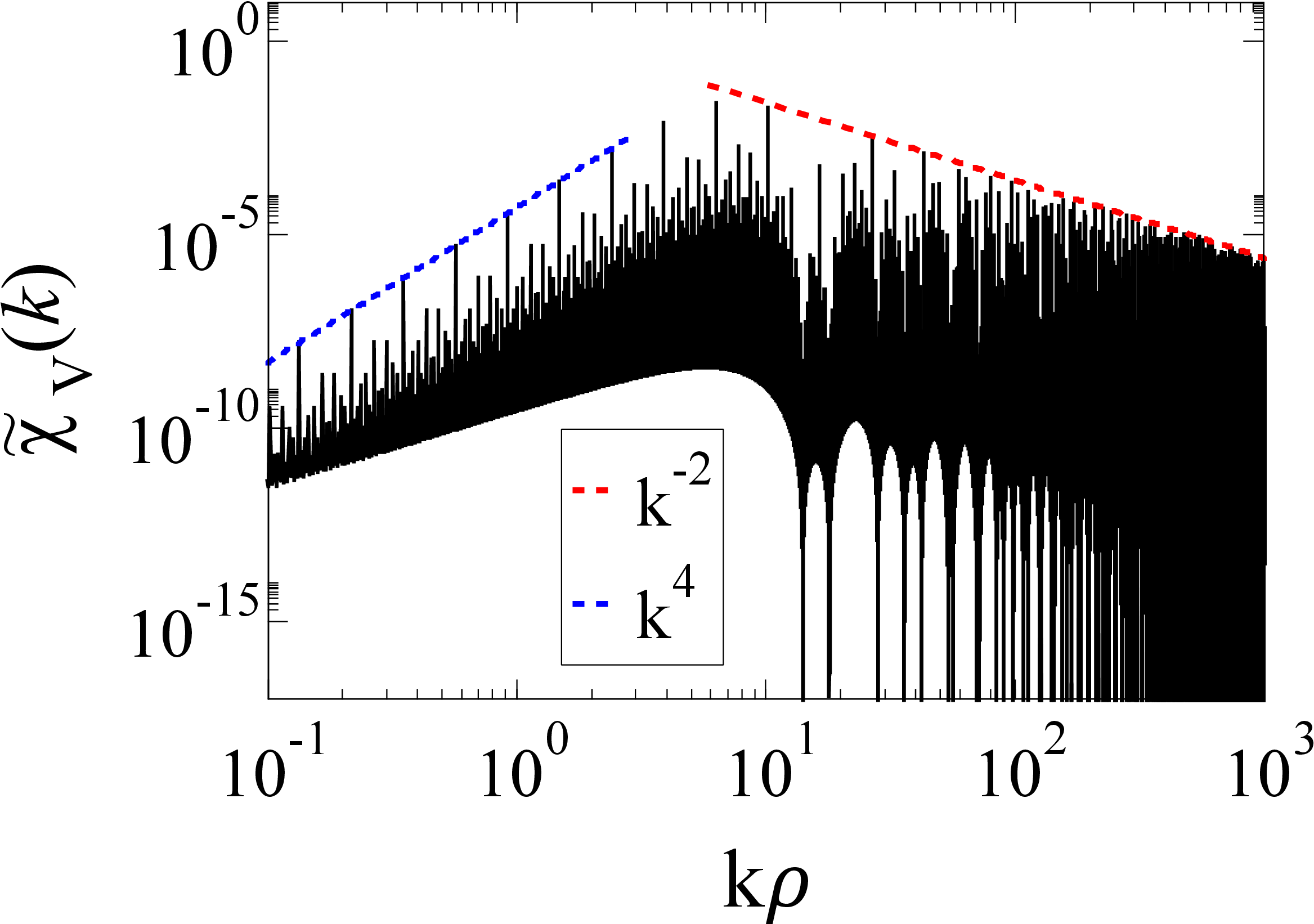} }\\
    \subfigure[]{\includegraphics[width=0.45\textwidth]{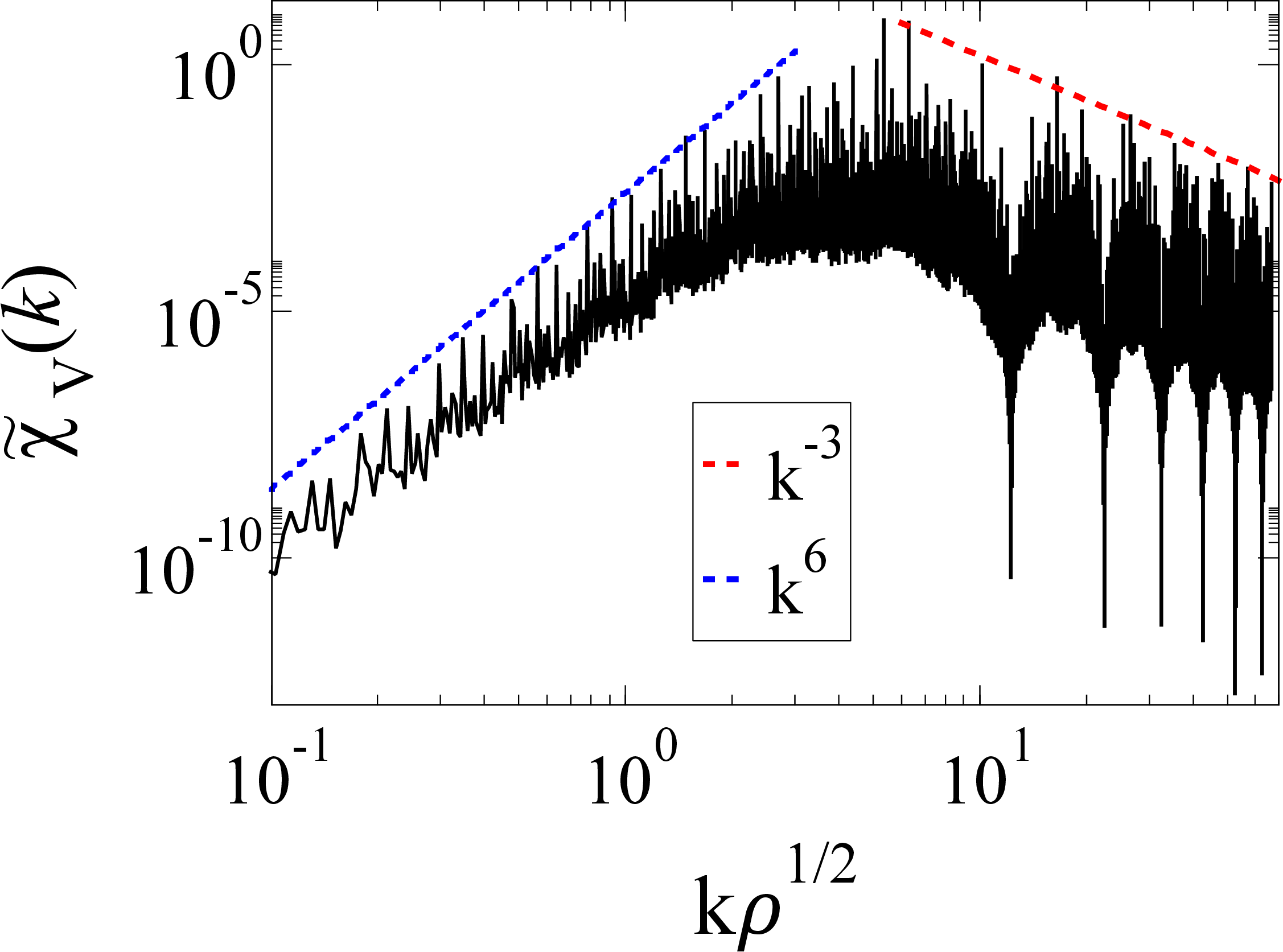} }
    \caption{A log-log plot of the spectral density $\Tilde{\chi}_{_V}(k)$ as a function of the scaled wavenumber $k\rho^{1/d}$ of a two-phase medium derived from (a) a Fibonacci chain with a packing fraction $\phi_2 = 0.35$, and (b) a Penrose tiling with a packing fraction $\phi_2=0.25$. The dashed red (gray) lines indicate (a) $k^{-2}$ scaling and (b) $k^{-3}$ scaling. The blue dashed lines indicate (a) $k^4$ scaling and (b) $k^6$ scaling.}
    \label{fig:largek_scale}
\end{figure}


\begin{thebibliography}{63}
\bibitem [{non()}]{nonan_foot}%
  \BibitemOpen
  \href@noop {} {}\bibinfo {note} {As described in Ref. \cite{hyperuniformstatesofmatter}, if $k^{\beta}$ is the lowest-order nonanalytic term in the spectral function (e.g., $\beta$ is odd or some real noninteger), then $\chi_{_V}(r) \sim 1/r^{d+\beta}$ for large $r$ in any space dimension $d$.}\BibitemShut {Stop}%
\end{thebibliography}

\providecommand{\noopsort}[1]{}\providecommand{\singleletter}[1]{#1}%

\end{document}